%% file: paper.tex
\pdfoutput=1
\documentclass[preprint,12pt]{elsarticle}

\input{preamble}

\usepackage{hyperref}
\usepackage{nnpdf}

\begin{document}

\begin{frontmatter}
  
  \title{
{\normalsize\noindent
\hfill MCnet-13-21\\[-1mm]
\hfill Edinburgh 2013/36}\\[5mm]
MCgrid: projecting cross section calculations on grids}

  \author[a]{Luigi Del Debbio}
  \author[a,b]{Nathan Hartland\corref{author}}
  \author[b]{Steffen Schumann}

  \address[a]{The Higgs Centre for Theoretical Physics, University of Edinburgh, UK.}
  \address[b]{II. Physikalisches Institut, Georg-August-Universit\"at G\"ottingen, G\"ottingen, Germany.}
  \cortext[author]{Corresponding author.\\\textit{E-mail address:} mcgrid@projects.hepforge.org}

  \begin{abstract}
    \packagename is a software package that provides access to the \appl interpolation tool for Monte Carlo event generator codes, allowing for fast and flexible variations of scales, coupling parameters and PDFs in cutting edge leading- and next-to-leading-order QCD calculations.  This is achieved by providing additional tools to the \rivet analysis system for the construction of \packagename enhanced \rivet analyses. The interface is based around a one-to-one correspondence between a \rivet histogram class and a wrapper for an \appl interpolation grid. The \rivet system provides all of the analysis tools required to project a Monte Carlo weight upon an observable bin, and the \packagename package provides the correct conversion of the event weight to an \appl fill call. \packagename has been tested and designed for use with the \Sherpa event generator, however as with \rivet the package is suitable for use with any code which can produce events in the {\tt HepMC} event record format.
  \end{abstract}


\end{frontmatter}

\noindent
{\bf PROGRAM SUMMARY}
\\
\begin{small}
  {\em Manuscript Title:} MCgrid: projecting cross section calculations on grids\\
  {\em Authors:} Luigi Del Debbio, Nathan Hartland, Steffen Schumann\\
  {\em Program Title:} MCgrid\\
  {\em Journal Reference: }                                      \\
  {\em Catalogue identifier:}                                   \\
  {\em Licensing provisions:} none                              \\
  {\em Programming language:} C++, shell, Python\\
  {\em Computer:} PC running Linux, Mac\\
  {\em Operating system:} Linux, Mac OS\\
  {\em RAM:} varying \\
  {\em Number of processors used:} 1                              \\
  {\em Supplementary material:}                                 
  {\em Classification:} 11.2 Phase Space and Event Simulation, 11.5 Quantum Chromodynamics, 11.9 Event Reconstruction and Data Analysis\\
  {\em External routines/libraries:} HepMC, Rivet, APPLgrid\\
  {\em Nature of problem:} Efficient filling of cross section 
  grid files from fully exclusive parton level Monte Carlo events\\
  \\
  {\em Solution method:} Analyse Monte Carlo events via the Rivet program, 
  which projects events on discretized cross section tables from APPLgrid\\
  \\
  {\em Running time:} varying.
  \\

\end{small}


\section{Introduction}
\label{sec:intro}
\input{intro}

\section{Parameter variation in NLO calculations}
\label{sec:theory}
\input{theory}

\section{The \packagename\hskip2pt Interface}
\label{part:interface}
\input{interface}

\section{Software Implementation}
\label{part:userguide}
\input{userguide}

\section{\packagename Validation}
\label{part:validation}
\input{validation}

\section{Conclusions}
\label{sec:conclusions}
\input{conclusions}

\appendix

\section*{Acknowledgements}
\label{app:acknowledgements}
\input{acknowledgements}



\bibliographystyle{elsarticle-num}
\bibliography{refs}

\end{document}

%% file: preamble.tex
\usepackage{xspace,graphicx,mparhack,amsmath}
\usepackage{amssymb,url,underscore,fancyvrb,cancel}
\usepackage{picinpar,fancybox}
\usepackage{microtype,relsize}
\usepackage{caption}
\usepackage{subcaption}

\usepackage{listings}
\usepackage{color}
\usepackage{textcomp}
\definecolor{listinggray}{gray}{0.9}
\definecolor{lbcolor}{rgb}{0.98,0.98,0.98}

\def\inline{\lstinline[basicstyle=\ttfamily,keywordstyle={}]}

\DeclareFontShape{OT1}{cmtt}{bx}{n}{<5><6><7><8><9><10><10.95><12><14.4><17.28><20.74><24.88>cmttb10}{}

\makeatletter
\g@addto@macro\bfseries{\boldmath}
\makeatother

\graphicspath{{./}{figs/}}
\newcommand{\packagename} {{\tt MCgrid }\!\!}
\newcommand{\rivet} {{\tt Rivet }\!\!}
\newcommand{\Rivet} {{\tt Rivet }\!\!}
\newcommand{\appl} {{\tt APPLgrid }\!\!}
\newcommand{\Sherpa} {{\tt SHERPA }\!\!}

\let\oldmarginpar\marginpar
\renewcommand\marginpar[1]{\-\oldmarginpar{\footnotesize \textit{#1}}}

\newcommand{\warnimg}{\includegraphics[height=11mm]{warning}}
\newcommand{\coneimg}{\includegraphics[height=11mm]{cone}}
\newcommand{\bendimg}{\includegraphics[height=11mm]{bend}}
\newcommand{\dblbendimg}{\bendimg\hspace{0.5mm}\bendimg}
\newcommand{\thinkimg}{\includegraphics[height=16mm]{thinker}}

\newcommand{\pT}{\ensuremath{p_\perp}\xspace}

%% file: intro.tex
Measurements at the Large Hadron Collider, operating at the highest centre-of-mass 
energies ever achieved in accelerator-based experiments, allow for precision 
studies of a vast range of final states. In particular large final state 
jet multiplicities become accessible. Such multi-jet events constitute both interesting 
signals and important backgrounds to new physics searches. Accordingly they must be 
described with a theoretical precision that either matches or exceeds the experimental accuracy. 

The dominant corrections to a given production process at a hadron collider originate from QCD. 
The demand for precise theoretical predictions describing LHC production processes has therefore 
triggered an enormous development in the field of next-to-leading-order QCD calculations -- making NLO QCD accuracy the new 
standard. These developments span from the largely automated calculation of NLO QCD 
cross sections to the combination of NLO matrix elements with parton-shower simulations, 
see e.g. Ref.~\cite{Buckley:2011ms} and references therein.  

The enormous progress recently experienced in the field of NLO QCD calculations
was sparked by two important developments. Firstly, the introduction of fast and efficient methods 
for the calculation of virtual amplitudes, see for instance~\cite{Denner:2005nn,Ossola:2006us,Anastasiou:2006jv,Ellis:2007br,Giele:2008ve,Berger:2008sj}. Secondly, the organisation and 
implementation of complete NLO calculations in the framework of parton level Monte-Carlo 
event generators such as {\tt Helac}~\cite{Bevilacqua:2011xh}, 
{\tt MadGraph}~\cite{Alwall:2011uj} or \Sherpa~\cite{Gleisberg:2003xi,Gleisberg:2008ta}. 
All of these approaches rely on using an infrared-subtraction formalism, e.g. the Catani--Seymour 
\cite{Catani:1996vz,Catani:2002hc} or the Frixione--Kunszt--Signer~\cite{Frixione:1995ms} 
method as implemented in automated subtraction term generators, see Refs.~\cite{Gleisberg:2007md,Czakon:2009ss,Hasegawa:2009tx,Frederix:2010cj,Bevilacqua:2013iha}. The real-emission 
corrections as well as the phase-space integration are handled by tree-level matrix-element 
generators such as {\tt AMEGIC}~\cite{Krauss:2001iv}, {\tt COMIX}~\cite{Gleisberg:2008fv}, 
{\tt MadGraph}~\cite{Alwall:2011uj} or {\tt Helac}~\cite{Cafarella:2007pc}. Virtual amplitudes, 
typically provided by specialised one-loop generators such as {\tt BlackHat}~\cite{Berger:2008sj}, 
{\tt GoSam}~\cite{Cullen:2011ac}, {\tt Helac-1Loop}~\cite{vanHameren:2009dr}, 
{\tt MadLoop}~\cite{Hirschi:2011pa}, {\tt NJET}~\cite{Badger:2012pg}, 
{\tt OpenLoops+Collier}~\cite{Cascioli:2011va} or {\tt Recola}~\cite{Actis:2012qn} can be 
incorporated via the BLHA interface~\cite{Binoth:2010xt,Alioli:2013nda}. Examples of recent 
NLO calculations that have been performed using a combination of the tools listed above 
include: $W+4,5$ jets \cite{Berger:2010zx,Bern:2013gka}, $Z+4$ jets \cite{Ita:2011wn}, 
$4-$jet and $5-$jet production \cite{Bern:2011ep,Badger:2012pf,Badger:2013yda}, 
$t\bar{t}+2$ jets \cite{Bevilacqua:2010ve} and $\gamma\gamma+2$ jets \cite{Gehrmann:2013bga}. 
Most of these new tools are now available to perform NLO QCD event generation and for use 
in LHC data analysis. 

In order to provide a comprehensive analysis of the 
uncertainty on an NLO QCD prediction, the value of the strong coupling, parton distribution functions,
renormalisation and factorisation scales must all be varied to characterise the dependence of the
final result upon these uncertain parameters. In particular, the accurate propagation of PDF
uncertainty through the calculation requires a great deal of repeated runs. 
For similar reasons, it is often challenging to include collider processes in global PDF 
fits beyond the LO approximation. Indeed, for cutting edge high-multiplicity processes the
frequent repetition of the full NLO calculation with a modified set of parameters is 
prohibitive as it requires too much CPU time.  

There are two complementary solutions to this problem that allow for an a-posteriori variation
of scale choices and parameters. Firstly, parton-level events supplemented with suitable 
weights can be stored explicitly. One such approach based on {\tt ROOT} NTuples has recently been presented in \cite{Bern:2013zja}. For the price of having large event files and having to loop over many events, one can in principle analyse any final state observable with arbitrary parameter choices for the process under consideration. The second approach is based on cross section interpolation grids that represent a given differential cross section binned in the incoming partonic momentum fractions $x$ and the associated process' scale $Q^2$. Examples of such interpolation tools are \appl \cite{Carli:2010rw} and {\tt FastNLO} \cite{Kluge:2006xs,Wobisch:2011ij}. While interpolating grids are defined only for a specific final state observable, they allow for a very quick re-evaluation of the respective cross section and have considerably reduced disk space requirements.

In this note we present \packagename\!, a tool that provides a direct interface from parton level Monte Carlo event generators performing LO or NLO QCD calculations to \appl\!. We utilise the \rivet\cite{Buckley:2010ar} MC analysis system to provide the experimental analysis tools needed in classifying the event final state into the appropriate observable bins. We convert the event weight information into an appropriate \appl fill call, while correctly taking into account the full PDF dependence. In the implementation we assume the event weights are generated via the Catani-Seymour dipole subtraction scheme.  Furthermore we provide the tools for an automated determination of the breakdown of a process into its contributing subprocesses. We can identify individual initial state flavour channels that can be combined, once their PDF dependence is factored out. \packagename is implemented as a {\tt C++} framework providing standard \rivet analyses with additional interpolation grid functionality. 

The present paper is organised as follows. In section \ref{sec:theory} we give a detailed account of the ingredients needed for a consistent parameter variation in QCD calculations, and we describe the basics of interpolation tools such as \appl\!. In section \ref{part:interface} we present our implementation of an interface allowing for the fill of \appl cross-section tables from fully exclusive parton-level Monte Carlo events in the {\tt HepMC} format.  In section \ref{part:userguide} we give detailed information on how to setup and execute \packagename\!.  We give examples and describe the validation of our approach in section \ref{part:validation}. The conclusions are presented in section~\ref{sec:conclusions}.

%% file: theory.tex
In this section we shall perform a brief overview of the methods and techniques available
for performing efficient variation of QCD parameters in NLO calculations.

\subsection{Reweighting leading order MC calculations}
\label{sec:paramvary}
QCD computations of final state observables involve detector acceptances, or jet algorithms, in
the computation of the perturbative coefficients. The latter are then convoluted with PDFs that
encode the nonperturbative information about the partonic content of hadrons.  Let us start by examining
the case of such a calculation at leading order in the strong coupling, computing a cross section for the production of some final state $X$,

\begin{equation}
  \sigma_{pp\to X}^{\text{LO}} = \int dx_1\, dx_2\; \int d\Phi_n \left( \frac{\alpha_s(\mu_R^2)}{2\pi} \right)^{p_{\text{LO}}} f_i(x_1,\mu_F^2) f_j(x_2,\mu_F^2) \, d\hat{\sigma}_{ij\to X}. \label{eq:LOconv}
\end{equation}

The PDFs for the partons inside the nucleons are denoted by $f_i$ and $f_j$, and the sum over
all partonic channels is understood. We have explicitly written the dependence of the PDFs on
the momentum fractions $x_1$, $x_2$ and on the factorisation scale $\mu_F^2$. The calculation's dependence on the value of the
strong coupling constant $\alpha_s$ is also made explicit. $d\hat{\sigma}_{ij\to X}$ is the parton-level squared matrix
element for the $2\to n$ process, differential in the final state phase space. 

This convolution may be simplified by making use of the initial state flavour symmetries of the parton level process $\hat{\sigma}$. Grouping the partonic sub channels which differ only by their PDFs into QCD subprocesses, Eq.~(\ref{eq:LOconv}) can be written as
\begin{equation}
    \sigma_{pp\to X}^{\text{LO}} = \int dx_1\, dx_2\; \int d\Phi_n \left( \frac{\alpha_s(\mu_R^2)}{2\pi} \right)^{p_{\text{LO}}} F_l(x_1,x_2,\mu_F^2) \, d\hat{\sigma}_{l\to X}\,,
  \label{eq:LOsubconv}
\end{equation}
where the subprocess parton density is given by
\begin{equation}
F_l(x_1,x_2,\mu_F^2) = \sum_{i,j=0}^{N_{pdf}}C_{ij}^{(l)} f_i(x_1,\mu_F^2)f_j(x_2,\mu_F^2)\,.
\end{equation}
The matrix of coefficients $C_{ij}^{(l)}$ is specified by the symmetries of the parton level cross section, determined by whether or not the partonic channel $ij$ belongs to the subprocess $l$, i.e.
\begin{equation*}
  C_{ij}^{(l)} = \quad
  \begin{cases}
    1, & ij \in l,\\
    0, & ij \notin l.
  \end{cases}
\end{equation*}

The integral in Eq.~(\ref{eq:LOsubconv}) can be computed by Monte Carlo integration: 
\begin{align}
  \label{eq:basicMC}
  \sigma_{pp\to X}^{\text{LO}} &= \sum_{e=1}^{N_\evt} \left(  \frac{\alpha_s\left(k_e\right)}{2\pi} \right)^{p_{\text{LO}}} w_e(k_e) F_{l_e}(k_e)\, ,
\end{align}
\begin{align}
  \label{eq:wedef}
  w_e(k_e) =d \hat{\sigma}_{l_e \to X}(k_e) \Pi_{\text{ps}}(k_e) \Theta (k_e-k_{\text{cuts}}),
\end{align}
where the index $e$ runs over the sample of generated MC events, $\Pi_{\text{ps}}(k_e)$ is the corresponding event phase space weight, $F_{l_e}$ is the subprocess density for the event subprocess $l_e$ and 
\begin{align}
  k_e &= \left\{p_1, ..., p_n,\,x_1,\,x_2,\,\frac{\mu_F^2}{Q^2},\,\frac{\mu_R^2}{Q^2} \right\},
\end{align}
are the set of kinematics associated with the event.
Eq.~(\ref{eq:wedef}) shows that the weight $w_e$ is obtained by evaluating the short-distance
cross-section $\hat{\sigma}$ for the kinematics generated for the given event $e$, taking the
kinematic cuts and final state phase space weight into account.

At this order, the procedure for performing a variation of the QCD parameters present in the calculation is fairly straightforward.
Provided that the full event record is stored for each entry in the sum in Eq.~(\ref{eq:basicMC}), a different PDF may be used simply by multiplying each weight by factors of $F_{l_e}^{(\text{new})}/F_{l_e}^{(\text{old})}$. A similar procedure may of course be used to reweight the LO power of
$\alpha_S$, and to vary the perturbative scales to which the calculation's only dependence is through $\alpha_S$ and the PDFs.

\subsection{Reweighting NLO event weights}
\label{sec:NLOvary}
Performing an event weight reweighting at NLO provides more of a challenge. The parton level cross section develops a dependence upon the
perturbative scales used in the calculation, and a subtraction mechanism must be employed in the Monte Carlo
integration to ensure the cancellation of singularities in the evaluated integrands.

The use of a subtraction algorithm makes the precise PDF and scale dependence of the event weight considerably
more complicated, due to the presence of integrated subtraction terms proportional to Altarelli-Parisi splitting functions. 
In this section we shall discuss the reweighting of events produced from the \Sherpa event generator which utilises
the Catani--Seymour dipole subtraction method~\cite{Catani:1996vz}. 

In general, a subtraction scheme will separate the differential cross section at NLO into four distinct parts.
\begin{equation}
  \label{eq:NLOxsect}
  \sigma_{pp\to X}^{\text{NLO}}= \int d\hat{\sigma}^\mathrm{B}
  + \int d\hat{\sigma}^\mathrm{V}
  + \int d\hat{\sigma}^\mathrm{I}
  + \int d\hat{\sigma}^\mathrm{RS}
  \, .
\end{equation}

When attempting to reweight an NLO event sample, the weights
must be treated differently according to whether they belong to 
the Born (B) or Real Subtracted (RS) integrals, or if they correspond to a Virtual (V)
or Integrated subtraction (I) event. 

In the case of the B or RS weights, their treatment is identical to
the leading order case, demonstrated in Eq.~(\ref{eq:basicMC}) and they may be reweighted by multiplication with the
appropriate new PDF and strong coupling factors. The B and RS integrals may be performed as

  \begin{align}
  \label{eq:Vcontrib}
   \int d\hat{\sigma}^\mathrm{B/RS}  &= \sum_{e=1}^{N_\evt} \left(  \frac{\alpha_s(\mu_R^2)}{2\pi} \right)^{p} F_{l_e}(k_e)\;  w^{(0)}_e(k_e),
\end{align}
where $p$ is either the leading (B) or next-to-leading (RS) order of $\alpha_S$ and we denote weight contributions with Born-like scale and PDF dependence by $w^{(0)}$.

For the events originating from virtual diagrams, the renormalisation of the 
matrix element introduces an explicit dependence on the renormalisation scale.
In order to accurately reweight the sample, the terms proportional to scale logarithms should
be kept track of separately.

  The Monte Carlo integral of the virtual contribution for an arbitrary choice of the renormalisation
  scale $\mu_R$ can then be computed as

  \begin{align}
   \int d\hat{\sigma}^\mathrm{V} &= \sum_{e=1}^{N_\evt} \left(  \frac{\alpha_s(\mu_R^2)}{2\pi} \right)^{p_{\text{NLO}}} F_{l_e}(k_e) \left\{ w^{(0)}_e(k_e) + \ell w^{(1)}_e(k_e) + \ell^2w^{(2)}_e(k_e)  \right\},
\end{align}
  where $\ell = \log\left(\frac{\mu_R^2}{\mu_{R, \text{old}}^2}\right)$. It is therefore clear that to provide an accurate scale variation the additional contributions $w^{(1)}$ and $w^{(2)}$ must be distinguished from the central scale weight $w$ in the event record.
  
  The third case, that of the integrated subtraction, introduces further complexity. 
  Specifically the PDF dependence of each event weight differs considerably from the leading order case.
  The counter terms introduced in the subtraction algorithm typically take the form of a 
  Born-type matrix element multiplied by a splitting function. These weights must therefore
  be expanded over many initial state parton flavours.
  
  To correctly perform a PDF reweighting then, the full dependence structure must also be detailed in the event record.
 If we assume that all integrated subtraction configurations are represented by just one event, the integral of the I contribution
 is performed as
\def\varref#1{{\tt #1}}
\begin{eqnarray} \label{eq:ISPDFdep}
 \int d\hat{\sigma}^\mathrm{I}  &=&  \sum_{e=1}^{N_\evt}   \left(  \frac{\alpha_s(\mu_R^2)}{2\pi} \right)^{p_{\text{NLO}}} \Bigg\{  f_1(i,x_1,\mu^2_F)\,  w^{(0)}_e \,f_2(j,x_2,\mu^2_F)
\nonumber\\
&&+\biggl(\sum_{k=1}^4 
   f_1^{(k)}(i,x_1,x'_1,\mu^2_F)\, w^{(3)}_{e,k}\biggr)\,
 f_2(j,x_2,\mu^2_F)\\
&&+f_1(j,x_1,\mu^2_F)
\biggl(\sum_{k=1}^4 
   w^{(4)}_{e,k}\, f_2^{(k)}(j,x_2,x'_2,\mu^2_F)\biggr) \Bigg\}
  \,, \nonumber
\end{eqnarray}
where here $x/x^\prime$ denote the values of parton-$x$ in the integration. The $w^{(3)}_k$ and $w^{(4)}_k$ denote the various contributions to the weight arising from the Altarelli-Parisi splitting functions for a parton splitting in the first beam or second beam respectively. Here we have used for the PDFs ($r=1$ or~$2$)
\begin{eqnarray}
f_r^{(1)}(i,x,x',\mu^2_F) &=& \left\{\begin{array}{l@{\hskip 0pt}cl}
  i = \mbox{quark} &:&f_r(i,x,\mu^2_F) \,, \\
  i = \mbox{gluon}&:&\sum\limits_{q} f_r(q,x,\mu^2_F)
   \vphantom{\displaystyle\sum} \,,
                 \end{array}\right.
\\
f_r^{(2)}(i,x,x',\mu^2_F) &=& \left\{\begin{array}{l@{\hskip 0pt}cl}
          i = \mbox{quark} &:&
                  f_r(i,x/x',\mu^2_F)/x' \,, \\
          i = \mbox{gluon}&:&\sum\limits_{q}\vphantom{\displaystyle\sum}
                  f_r(q,x/x',\mu^2_F)/x' \,,
              \end{array}\right.\\
f_r^{(3)}(i,x,x',\mu^2_F) &=& f_r(g,x,\mu^2_F) \,, \\
f_r^{(4)}(i,x,x',\mu^2_F)&=& f_r(g,x/x',\mu^2_F)/x' \,.
\end{eqnarray}
The sums over quarks $q$ are taken over the active quark/anti-quark flavours at scale $\mu^2_F$. Furthermore, $i,j$ specify the incoming parton flavours in the event $e$ before any splitting, with momentum fractions $x_1$ and $x_2$, respectively.\\\\
Projecting each individual partonic channel weight onto the subprocess basis, the same integral can be done as
\def\varref#1{{\tt #1}}
\begin{eqnarray}
 \int d\hat{\sigma}^\mathrm{I}  &=&  \sum_{e=1}^{N_\evt}   \left(  \frac{\alpha_s(\mu_R^2)}{2\pi} \right)^{p_{\text{NLO}}} \Bigg\{  F_{l_e}(x_1,x_2,\mu^2_F)\,  w^{(0)}_e 
\nonumber\\
&&+\sum_{s}^{N_{\text{sub}}} F_s(x_1/x'_1, x_2, \mu^2_F)\, \tilde{w}^{(3)}_{e,s} \\
&&+\sum_{s}^{N_{\text{sub}}} F_s(x_1, x_2/x'_2, \mu^2_F)\, \tilde{w}^{(4)}_{e,s}
 \Bigg\}
  \,, \nonumber
\end{eqnarray}
where the subprocess basis weights $\tilde{w}^{(3/4)}_s$ are obtained from the flavour basis weights $w^{(3/4)}_k$.
To be able to perform the PDF reweighting, the values of the additional weights and $x^\prime$ values
must be made explicit in the event record. 

In summary, there are five classes of contributions that should be distinguished in each event weight.
\begin{itemize}
\item $w^{(0)}$: Weights with Born-like PDF and scale dependence.
\item $w^{(1)}$: Weights proportional to first order scale logs.
\item $w^{(2)}$: Weights proportional to second order scale logs.
\item $w^{(3)}$: Weights originating from integration over $x_1$ in I events.
\item $w^{(4)}$: Weights originating from integration over $x_2$ in I events.
\end{itemize}
All of which must be separated into independent initial state flavour contributions, and possibly projected onto a
subprocess basis. It can be seen however, that with a central scale choice (i.e neglecting the $w^{(1/2)}$ terms)
it is possible to convert the full NLO calculation into a form similar to the leading order case in Eq.~(\ref{eq:basicMC}).
By projecting the $w^{(3/4)}$ weights onto independent events, we may write the full NLO result at central scales as
\begin{align}
  \label{eq:basicNLOMC}
  \sigma^\mathrm{NLO}_{pp\to X} &= \sum_{e=1}^{N_\evt} \left(  \frac{\alpha_s\left(k_e\right)}{2\pi} \right)^{p_{e}} F_{l_e}(k_e)\; w_e(k_e) ,
\end{align}
where $p_e$ denotes the order in $\alpha_S$ of the event $e$.

A prescription for identifying all these contributions in NLO records was recently described,
along with a procedure and software package for performing the reweighting in the {\tt ROOT}
NTuple\cite{Bern:2013zja} format designed by the {\tt BlackHat} collaboration and implemented in {\tt
  SHERPA}. With the various complexities of reweighting NLO events in the Catani--Seymour formalism
  carefully treated as in the {\tt BlackHat} NTuple format, all the required information is present to perform an accurate variation of the QCD 
parameters present in the calculation.

However there remains a difficulty with the event reweighting approach, in that the
full event record must be stored and reanalysed for each variation of a parameter in the
calculation. In high statistics samples, this can mean storing hundreds of gigabytes of event
files. Reanalysing these events may still be a nontrivial computational task simply in the
reading and reprocessing of events. Certainly for performance sensitive applications such as
PDF fitting, a reweighting approach remains prohibitively expensive.

\subsection{Interpolation tools} 
\label{sec:interp}

The fundamental difficulty in the event reweighting procedure lies in the sum over events in
Eq.~(\ref{eq:basicNLOMC}). In applications where both speed and accuracy are important, such a
dependence on the statistics of the event sample is problematic. Possible solutions to this
problem have been available for some time in the form of event weight interpolating tools such
as the \appl~\cite{Carli:2010rw} and {\tt FastNLO}~\cite{Kluge:2006xs,Wobisch:2011ij}
projects. The basic principle of these methods is to represent the PDFs used in the product in
Eq.~(\ref{eq:basicNLOMC}) upon an interpolating grid in $x$ and $Q^2$. 

The number of
points in each direction is denoted by $N_x$ and $N_Q$ respectively; therefore:
\begin{align}
  F_l(x_1,x_2,Q^2) &= \sum_{\alpha\beta}^{N_x} \sum_{\tau}^{N_Q} F_l(x_\alpha,x_\beta,Q_\tau^2)\,
  \mathcal{I}^{(\alpha)}(x_1) \mathcal{I}^{(\beta)}(x_2)   \mathcal{I}^{(\tau)}(Q^2)\,, \\
  &= \sum_{\alpha\beta}^{N_x} \sum_{\tau}^{N_Q} F_l(x_\alpha,x_\beta,Q_\tau^2)\, \mathcal{I}^{\alpha\beta,\tau}(x_1,x_2,Q^2).
\end{align}
If we once again neglect at first the terms in the calculation proportional to logarithms of the chosen scales, the Monte Carlo computation of the cross section in Eq.~(\ref{eq:basicNLOMC}) can be written
as:
\begin{align}
  \sigma &= \sum_e \sum_{\alpha\beta}^{N_x} \sum_{\tau}^{N_Q} \; \left(  \frac{\alpha_s\left(Q^2_\tau\right)}{2\pi} \right)^{p_{e}} 
  F_{l_e} (x_\alpha,x_\beta, Q_\tau^2)\; \mathcal{I}^{\alpha\beta,\tau}(k_e) \; w_e(k_e) .
  \end{align}
Subdividing the weights by perturbative order and rearranging the terms in the sum yields:
\begin{align}
  \label{eq:scalprod}
  \sigma &= \sum_{p}\sum_{l}^{N_{\text{sub}}} \; \sum_{\alpha\beta}^{N_x} \; \sum_{\tau}^{N_{Q}}\; 
  \left(  \frac{\alpha_s\left(Q^2_\tau\right)}{2\pi} \right)^{p}\;
  F_l(x_\alpha,x_\beta, Q^2_\tau) \; W^{(l)(p)}_{\alpha\beta,\tau}\, ,
\end{align}
where the sum over the events is performed: 
\begin{align}
   \label{eq:weightdef}
  W^{(l)(p)}_{\alpha\beta,\tau} &= \sum_e \left[
    \delta_{l, l_e} \delta_{p, p_e}\:
    \mathcal{I}^{\alpha\beta,\tau}(k_e)\;  w_e(k_e)
    \right].
\end{align}
Note that Eq.~(\ref{eq:scalprod}) is a simple sum over the points in the $x$, $Q^2$ grids. These kinematic points are chosen when defining
the grids, and do not change. Once the interpolated weights $W^{(l)(p)}_{\alpha\beta,\tau}$ for a given
process are stored, the computation of the cross section is very fast, since it does not
involve a loop over the generated MC events.

If we are interested in performing a calculation differential in some observable or otherwise a quantity that may be represented
in a histogram, the event weight final states quantified in the kinematics $k_e$ should be projected onto the relevant observable bin.
In this way, the interpolated weight grid is separated into final state observable bins,
\begin{align}
   \label{eq:weightdefbinned}
  W^{(l)(p)(b)}_{\alpha\beta,\tau} &= \sum_e \left[
    \delta_{l, l_e} \delta_{p, p_e}\:
    \mathcal{I}^{\alpha\beta,\tau}(k_e)\;  w_e(k_e) \; \Theta(k^{\text{max}}_b - k_e) \Theta(k_e - k_b^{\text{min}})
    \right]\,,
\end{align}
where the $k_b^{\text{min}}$, $k_b^{\text{max}}$ represent the required kinematic limits for the observable bin $b$. The cross section calculation in Eq.~(\ref{eq:scalprod}) is therefore
separated into a differential observable. As the sum over the events has been performed,
there is no requirement for the detailed event by event information present in the full record, drastically reducing
the storage space requirements. Of course this comes with a caveat. Once the weights are interpolated and
stored, the produced grid is restricted to the experimental projection defined by the cuts used when constructing it. As the event
final state kinematics are discarded after the projection, each grid is uniquely identified by its projection and binning.

In such an approach the additional scale dependent terms described in the previous section are discarded. Therefore it
may at first seem that performing scale variations would be challenging. However, these terms may be
inferred by a simple calculation that is made considerably easier by the presence of the full interpolated weight sample,
separated by perturbative order.

By demanding that the scale derivative of the cross section result be zero to $\mathcal{O}(\alpha_S^{\text{NNLO}})$
the terms proportional to logarithms of the factorisation and renormalisation scales may be calculated. This procedure
is discussed in detail in Ref. \cite{Carli:2010rw}.

%% file: interface.tex
Access to interpolating tools has so far been restricted to NLO cross section calculators such as {\tt MCFM} \cite{mcfm} and {\tt nlojet++ }\cite{Nagy:2003tz}. Here we shall describe an interface, named \packagename\!, that will allow for efficient access to the latest NLO calculations as obtained with fully exclusive parton level event generators.

We are using the \appl framework to provide the interpolation grid definitions, along with its comprehensive suite of tools for recalculating the stored cross section with varied parameters and scales. The \appl package provides a set of standard methods for the construction and filling of the interpolated weight grids described in section \ref{sec:interp}.

In order to develop an interface for event generators to the \appl package, a number of issues should be addressed. Firstly the projection of each event weight upon an observable bin, quantified in Eq.~(\ref{eq:weightdefbinned}), must be performed by some set of analysis tools.

Secondly, the PDF dependence of each event weight from a generator may be complicated by the presence of integrated subtraction terms in the event sample as demonstrated in section \ref{sec:NLOvary}. Such complex PDF dependence should be removed by converting the single event weight into individual fills, each corresponding to a single pair of initial state partons.

Finally, full Monte Carlo event generators typically differ from NLO cross section integrators in their evaluation of the cross section. While many NLO codes will perform the integration directly in the subprocess PDF basis as in Eq.~(\ref{eq:LOsubconv}), event generators will generally produce weights exclusive in the initial state, meaning that the integral in Eq.~(\ref{eq:LOconv}) is performed with the full PDF basis. As the full basis is rather inefficient for the reweighting of a fixed-order calculation, the produced parton-parton weights should be converted into subprocess weights while preserving the statistical accuracy of the calculation. 

The \packagename package provides a conversion of NLO event generator weights into a form suitable for interpolation. The projection of each event weight onto an observable bin is performed by the \Rivet analysis system. In this section we shall briefly describe the features of the interface along with a description of how it may be used in practice.

\subsection{Event final state analysis}
When developing an interface to process Monte Carlo events into observable bins in an interpolated weight grid, clearly a suite of experimental analysis tools are required in order to perform the projection from the event's final state to the appropriate observable bin. Rather than re-implement such an analysis suite we have opted to make use of the flexibility of the \rivet analysis system to provide such tools.

The \rivet framework provides a standard set of analysis methods, along with tools for the reading of event records from disk or processed on the fly directly from an event generator. The \rivet system is becoming a standard in Monte Carlo analysis for LHC observables. While it is geared mainly for the analysis of parton showered/hadronized final states, it can also process events from fixed order calculations. A typical \rivet analysis is associated with an experimental measurement, and as such generally includes the experimental data and  binning information to provide an automated comparison. However, the inclusion of experimental reference data is not a requirement. 

The \packagename interface provides a set of additional methods to an analysis procedure in \Rivet for the generation of \appl interpolated weight grid files. The final state cuts and binnings are performed as usual in a \Rivet analysis, with the \packagename functionality requiring only simple modifications to produce a weight grid. The interface has been designed to follow \Rivet standards as much as possible, with each produced \appl file corresponding to a histogram in the analysis. 

\subsection{Interpolating NLO event records}
\label{sec:IFInterNLO}

\rivet analyses are based upon event records in the standard {\tt HepMC} format \cite{Dobbs:2001ck}. By design this format provides mostly final state  information, therefore some additional data is required in order to fill the \appl weight grids.

The information on the parton distribution functions and the corresponding $x$ and $Q^2$ values are already hosted by the {\tt HepMC::PDFInfo} class. The running coupling evaluated at the process' renormalisation scale is provided via the method {\tt HepMC::GenEvent::alphaQCD()}. This information should be correctly filled in the exported event record.

On top of this standard information, all {\tt HepMC} events that are passed to an \packagename enabled analysis must also include the power of the strong coupling characterising each event. This is required so that the correct power may be removed from the event weight, it must therefore be the absolute power rather than the power relative to leading order. 

Additionally in the filling of an interpolation grid from an NLO event weight the issue of the precise PDF dependence of the weight arises as described in Section \ref{sec:paramvary}. In \packagename we assume by default that each integrated subtraction event weight carries this full PDF dependence. The full event weight must therefore be expanded in a basis of terms corresponding to each application of a splitting function, depending on the factorisation scheme used to define the PDF set. The \packagename interface expects as inputs the same basis as is used for the PDF reweighting terms in the {\tt BlackHat} NTuple format described in \cite{Bern:2013zja} with the exception of the additional weights accounting for the possible variation of the factorisation scale which are not needed when performing scale variation using the approach described in Ref. \cite{Carli:2010rw}.

This non-standard information should be appended to each event in the record as additional entries in the {\tt HepMC::WeightContainer} vector. In this way, no modification to the standard {\tt HepMC} format is necessary. Explicitly, the full set of {\tt HepMC::WeightContainer} entries expected by \packagename is:
\begin{itemize}
\item {\tt{WeightContainer[0-3]}} = [Reserved for Generator use]
\item {\tt{WeightContainer[4]}} = Event's power of $\alpha_S$
\item {\tt{WeightContainer[5]}} = Total event weight $\propto f_{i/H1}f_{j/H2}$\,, \\
where $ij$ labels the hard event's initial state flavours.
\item {\tt{WeightContainer[6]}} = Number of additional weights to come.
\item {\tt{WeightContainer[7]}} = Secondary momentum fraction $x^\prime_1$ for beam 1
\item {\tt{WeightContainer[8]}} = Secondary momentum fraction $x^\prime_2$ for beam 2
\item {\tt{WeightContainer[9+i]}} = {\tt{usr\_wgts[2+i]}}\,,\\
                                      where $i=0,\dots,7$ and with the {\tt usr\_wgts} defined according to the basis in \cite{Bern:2013zja}\,.
\end{itemize}
Here in the {\tt HepMC::WeightContainer} all powers of $\alpha_S$ should remain present in the specified weights, but the PDF values should not, as is the case for the {\tt{BlackHat}} NTuple record. It is worthwhile noting here that for Born-like events, real-emission corrections and real-emission counter-configurations no additional weight information needs to be provided, it suffices to have the {\tt{HepMC::PDFInfo}} properly filled, along with the standard event weight and its power of the strong coupling. In this instance the {\tt HepMC::WeightContainer} entry 6 should be set to zero.

The additional weight information listed above is needed, as we assume that all integrated subtraction configurations get represented by just one event. Accordingly the corresponding event weight carries a dependence on various initial state flavour combinations and PDFs. For a consistent filling of cross section grids we have to disentangle all these contributions and fill the partial event weights in the suitable subprocess grid. Recalling the decomposition of the full event weight from Ref.~\cite{Bern:2013zja} and described in Eq.~(\ref{eq:ISPDFdep}) the decomposition of the full event weight $w$, allowing for a consistent variations of PDFs, reads
\def\varref#1{{\tt #1}}
\begin{eqnarray}
w &=& \varref{WeightContainer[5]}
\nonumber\\
&&+\biggl(\sum_{k=1}^4 
   f_1^{(k)}(i,x_1,x'_1,\mu^2_F)\, w_k\biggr)\,
 f_2(j,x_2,\mu^2_F)\\
&&+f_1(j,x_1,\mu^2_F)
\biggl(\sum_{k=1}^4 
   f_2^{(k)}(j,x_2,x'_2,\mu^2_F)\, w_{k+4}\biggr)
  \,, \nonumber
\end{eqnarray}
where the decomposed weights and $x^\prime$ values must be provided by 
\begin{eqnarray}
x'_1 &=& \varref{WeightContainer[7]}\,,\nonumber\\
x'_2 &=& \varref{WeightContainer[8]}\,,\nonumber\\
w_k &=& \varref{WeightContainer[}k\varref{+8]} \,.\nonumber
\end{eqnarray}

In the case where the Monte Carlo code you wish to interface with \packagename provides also the integrated counter-term weights as independent events, the full PDF dependence of the event sample can be described by Eq.~(\ref{eq:basicNLOMC}). If this is the case, then all event weights may be treated as in the Born or real-emission cases, with the {\tt HepMC::WeightContainer} entry 6 zeroed.

These modifications have been implemented into the {\tt HepMC_Short} output of \Sherpa as of version 2.0. Therefore \packagename is able to process \Sherpa output without modification. 

\subsection{Automated subprocess determination}
\label{sec:subprocdet}
Assuming fully exclusive parton level events as inputs for \packagename we have to deal with fully exclusive partonic initial states. This corresponds to the maximal dimensionality of the flavour basis for a considered process. Considering proton-proton collisions with $\{u,d,s,c,b,g\}$ and the corresponding anti-quarks as initial state partons we can in principle have 121 different partonic initial states.  Obviously for each process there exists an often \emph{much smaller} basis of distinct subprocesses that combine individual channels that are identical up to the initial state PDFs, cf. Sec.~\ref{sec:interp}.

In previous \appl applications these subprocess bases have been identified manually, although a method for automated subprocess identification was sketched in \cite{Carli:2010rw}. Here we propose to use process information that is readily available from the event generators themselves, here in particular \Sherpa\!. 

At the level of the generation of the matrix elements for a given process individual channels get mapped onto each other. This is achieved by a one-to-one comparison of the transition amplitudes. This procedure guarantees an efficient re-use of matrix element expressions and significantly speeds-up cross section calculations. To give a simple example, the $c\bar{c}$ initiated contribution to inclusive QCD jet production in $pp$ or $p\bar p$ collisions is described by the very same matrix elements that account for the $u\bar{u}$ channel. As QCD interactions are flavour-diagonal the difference between these two channels originates from the initial state PDFs only.

We want to use this matrix element generator internal identification and mapping of equivalent 
partonic channels to determine the reduced subprocess flavour basis. For \Sherpa\!'s matrix element generators {\tt COMIX} and {\tt AMEGIC} the information on mapped channels is written out when the processes get generated for the first time. With \packagename we now supply python scripts to directly convert these process maps into the {\tt lumi_pdf} format used by \appl\!. For that purpose the scripts analyse all channel maps and effectively collect a table of distinct subprocesses with all their contributing flavour channels. While these process maps are generated from \Sherpa output, their validity is not limited to \Sherpa\!, rather they can be used with every other event generator delivering events fully exclusive in the initial state flavours.

\subsubsection{Statistical issues in subprocess identification}
\label{sec:subprocstat}
It should be noted that there are a number of subtleties involved when using the subprocess identification as described in \ref{sec:interp}. When filling grids with the subprocess identification enabled the cross section calculation applies the weights in the event sample to all the equivalent mapped channels in the same subprocess. As each subprocess can make use of more event weights than any of its component channels, the resulting total cross section enjoys an improved statistical accuracy, albeit mostly benefiting the partonic channels that weigh less in the total calculation. As an explicit example, compare the Monte Carlo sum in the two cases:

\begin{equation}
\sigma = \sum_{e} w_e(i,j) \; f_i \, f_j, \quad \quad \sigma_{\text{sub}}=\sum_{e} \sum_{l} w_e(i,j) \; C_{ij}^{(l)} F_l.
\end{equation}
While both estimators for the cross section are physically well motivated, it is clear that they differ when operating on a finite event sample. This difference makes a direct comparison of the subprocess-grouped result to the benchmark calculation (where no such grouping is performed) into a statistical exercise, with differences naturally tending to zero in the limit of very large samples.

In addition to the (slight) statistical advantage available when using a subprocess grouping, the resulting \appl files tend to be considerably smaller, as the number of contributing subprocesses for a particular reaction is typically much smaller than 121. This also impacts the final convolution time when making use of the produced \appl files. Indeed, to be competitive in applications such as PDF fitting, such a grouping is almost a necessity.

Secondly, in exclusive event generation channels typically get selected with a weight proportional to their relative contribution to the total cross section. In this way a better convergence of the cross section estimate using a finite number of phase space points is achieved. These selection weights need to be accounted for by the respective event weights. Accordingly, the event weights for rather rare processes are significantly enhanced. To be precise, the sampled partonic cross section in Eq.~(\ref{eq:wedef}) is complemented by a normalisation factor,
\begin{equation}  
w_e(i,j,k_e) =\mathcal{N}_{ij}\;d \hat{\sigma}_{l_e \to X}(k_e) \Pi_{\text{ps}}(k_e) \Theta (k_e-k_{\text{cuts}}),
\end{equation}
where the normalisation factor 
\begin{equation}
\mathcal{N}_{ij} \sim \frac{N_{\text{tot}}}{N_{ij}}
\end{equation}
can be accurately approximated by the ratio of the total number of events to the number in the $ij$ channel.

These selection weights render a naive grouping of channels into subprocesses very inefficient, as channels with poor statistics but comparably large weights would dominate the statistically uncertainty of even a well populated subprocess. To avoid this, we need to determine the relative population of all partonic channels in the process to account for the selection weights. In this way, a channel specific event may be converted into a subprocess event while preserving the statistical accuracy of the overall calculation, i.e.
\begin{equation}
  W^{(l)(p)(b)}_{\alpha\beta,\tau} = \frac{\mathcal{N}_{l}}{\mathcal{N}_{ij}}C_{ij}^{(l)}W^{(ij)(p)(b)}_{\alpha\beta,\tau}.
\end{equation}
The ratio $\mathcal{N}_{l}/\mathcal{N}_{ij}$ is determined numerically by monitoring the number of events falling into sub channel $ij$ relative to subprocess $l$.
In \packagename this is done in an initial loop over events, that must anyway be performed in order to determine phase space boundaries for the cross section grids, cf. Sec.~\ref{part:validation}.

Let us end by noting, that it is always possible to use the full flavour basis of 121 initial state combinations. In fact, it is only in this basis that we can guarantee to reproduce the input cross section within interpolation accuracy. However, the resulting grids will be significantly larger than corresponding grids produced with subprocess grouping enabled. In particular for applications such as PDF fitting subprocess identification is a must and with \packagename we provide very efficient and fully automated methods to accomplish this for arbitrary processes.

%% file: userguide.tex
In this section we shall briefly describe how the \packagename tool may be implemented into a \rivet analysis in practice. Here a typical implementation will be broadly sketched, for the detailed technical documentation please refer to the user manual included in the package.

The \packagename package is supplied as an external library which may be linked to a \rivet analysis at compile time. \packagename may be configured and installed in the conventional way with the autotools build system. The package may be configured for two main fill modes, the default fill behaviour takes into account the PDF structure of event weights originating from the \Sherpa event generator, cf. Section~\ref{sec:IFInterNLO}. For a generic fill mode where the PDF dependence is described fully by Eq.~(\ref{eq:basicNLOMC}), an option is available in the configuration. 

There are three main objects that are made available when linking an analysis to \packagename\!:
\begin{itemize}
\item \inline{MCgrid::mcgrid_pdf}\\
This object inherits from the \appl {\tt lumi_pdf} class. In addition to the subprocess identification, it provides the required subprocess event counting functionality as described in Section \ref{sec:subprocdet}. Initialised by \inline{MCgrid::bookPDF}.
\item \inline{MCgrid::PDFHandler}\\
This object is used to keep track of the initialised subprocess PDFs, and to pass events to them in the counting phase. Initialised at first use.
\item \inline{MCgrid::gridPtr}\\
The primary object in the package, this provides a wrapper for an \inline{APPL::grid} object. The class performs the conversion
of an event generator weight to a suitable \appl fill call. Provided as a\\ \inline{boost:smart_ptr} analogously to the \rivet histogram implementation. Initialised by \inline{MCgrid::bookGrid}.
\end{itemize}
We shall now summarise the modifications required to each analysis phase in order to produce an \appl file from \packagename\!.

\subsection{Initialisation Phase}

Initialising the \packagename tools in an analysis is a matter of booking the subprocess PDF descriptions for the process in question, and 
allocating the interpolation grid classes.  

This subprocess information is provided by \appl {\tt lumi_pdf} config files. For the details of how these files may be obtained from \Sherpa or constructed by hand, refer to the documentation supplied with \packagename\!.
To initialise a subprocess config file in \packagename the following method should be called in the initialisation phase for each process in the analysis:

\begin{lstlisting}[language=c++]
 MCgrid::bookPDF(configname, histoDir(), beam1Type, beam2Type);
\end{lstlisting}

Where \inline{configname} is an \inline{std::string} providing the filename of the subprocess config name. \inline{histoDir()} is a standard \rivet function which provides the name of the analysis. \inline{beam1Type} and \inline{beam2Type} specify whether the beam types used in the config file should be charge conjugated when performing a fill. This accounts for changing quarks to anti-quarks and vice versa in the case of an anti-proton beam. 

With the subprocess PDFs initialised it is time to set up the interpolating grids themselves. Firstly the \rivet analysis should be implemented and checked as in a standard analysis using only the histogram classes. Once the user is satisfied with the analysis, they should add to the analysis class
their grid classes with the following method:
\newpage
 
  \begin{lstlisting}[language=c++]
	MCgrid::gridPtr MCgrid::bookGrid( 
		// Corresponding Rivet histogram
		const Rivet::Histo1DPtr hist,      
		// Result of Rivet histoDir() call
		const std::string histoDir,        
		// APPLgrid subprocess PDF   
		const std::string pdfname,    
		// Leading order power of alpha_s for the process
		const int    LOpower,               
		// Minimum value of parton x in the event sample      
		const double xmin,                   
		// Maximum value of parton x in the event sample
		const double xmax,        
		// Minimum event scale^2           
		const double q2min,          
		// Maximum event scale^2       
		const double q2max,            
		// Grid architecture    
		const gridArch arch   
	);   
\end{lstlisting}

Where the struct \inline{gridArch} specifies the architecture of the \appl interpolation. It can be initialised with the following constructor:

  \begin{lstlisting}[language=c++]
	gridArch( 
		const int nX, 	// Number of points in x-grid
		const int nQ2,	// Number of points in Q^2 grid
		const int xOrd,	// Order of interpolation on x-grid
		const int Q2Ord // Order of interpolation on Q^2-grid
	):
 \end{lstlisting} 
 
 As an example \inline{init()} phase, consider the construction of a grid for a Drell-Yan $Z$-rapidity analysis where events are generated with a fixed scale of $M_Z^2$ from $p\bar{p}$ beams:
\newpage

   \begin{lstlisting}[language=c++]
   // Subprocess PDF
    const string PDFname("DYppbar.config");
    MCgrid::bookPDF(PDFname, histoDir(),
                 MCgrid::BEAM_PROTON, MCgrid::BEAM_ANTIPROTON);
                 
    // Grid architecture
    MCgrid::gridArch arch(50,1,5,0);
   
    /// Book histograms and grids
    _h_xsection = bookHisto1D(1, 1, 1);
    _a_xsection = MCgrid::bookGrid(	_h_xsection,
    								histoDir(), PDFname,
     								0, 
									1E-5, 1,
									8315.18, 8315.18,
									arch);
\end{lstlisting}

\subsection{Analysis phase}

In the analysis phase of the code, the first required modification is that the \packagename event handler must be called for every event passed to \rivet\!. This is done by adding the following line to the very start of the analysis phase, before any selection cuts:
\begin{lstlisting}[language=c++]
	MCgrid::PDFHandler::HandleEvent(event);
\end{lstlisting}
Once the events have been counted, both the histograms and \appl classes must be populated after the experimental cuts and analysis tools are applied as usual. Once the required event selection has been performed and the user is ready to fill a histogram, they simply have to fill the corresponding \inline{gridPtr} also: 
   \begin{lstlisting}[language=c++]
	_h_distribution->fill(coord, weight);	// Histogram fill  
	_a_distribution->fill(coord, event);	// Grid fill
\end{lstlisting}
Here \inline{coord} specifies the value of the binned quantity for that event, \inline{weight} is the usual event weight and \inline{event} is the \inline{Rivet::Event} object passed to the \inline{analyse} method.

\subsubsection{Finalise phase}
Finally the normalisation of the grids should be set, and the \appl \inline{.root} files exported for use. This is accomplished in the \inline{finalise} phase of the analysis. For the normalisation the treatment of the grids is once again analogous to that of the histograms. For each histogram/grid pair to be normalised the following should be called:
   \begin{lstlisting}[language=c++]
	// Histogram normalisation
	scale(_h_distribution, normalisation);	
	// Grid normalisation
	_a_distribution->scale(normalisation);	
\end{lstlisting}

And finally the grids should be written to file:   
\begin{lstlisting}[language=c++]
	_a_distribution->exportgrid();	
\end{lstlisting}
The filename of the grid will be based automatically upon the id of the corresponding histogram.

As the last modification step in the finalise phase, the event handler must be cleared and exported by adding the following as the final line in the finalise phase:
\begin{lstlisting}[language=c++]
	MCgrid::PDFHandler::ClearHandler();
\end{lstlisting}

\subsection{Executing a \packagename / \rivet analysis}
As is typical with the \appl package, to fill its produced grids two runs of the analysis must be performed. The first, or phase space fill run, determines the relative statistics of each partonic channel in the process such that their statistical samples may be combined correctly, and also establishes the boundaries of the $x$, $Q^2$ phase space for each of the interpolation grids as explained in \cite{Carli:2010rw}. The second run actually populates the grids with the Monte Carlo weights. It is therefore typically sufficient to perform a run with a smaller but representative event sample for the phase space run, and only run the full event sample for the full fill.
The modified \rivet analysis produced with \packagename utilities can be used as a completely conventional \rivet analysis, running over {\tt HepMC} event record files, or indeed streamed via a {\tt FIFO} pipe or straight from an event generator. Once the event sample has been run through the modified analysis twice, a standard \appl file will be produced.

%% file: validation.tex
In order to validate the interface and describe some details and options available when implementing an analysis in \rivet and \packagename we shall describe an application of the \packagename tool to two processes at hadron colliders; inclusive jet production and the Drell-Yan production of $Z$ bosons.

In the first part of this section, we shall examine the results in the full 121 subprocess basis. In this way a direct comparison to the benchmark cross section is possible. In the second part we examine directly the production of grids with subprocess grouping enabled.
\subsection{Interpolation accuracy and the ATLAS inclusive jet measurement}
The first test to validate the interface must of course test the ability of \packagename to generate a grid file that reproduces the benchmark result from the direct analysis of the events with \rivet\!. 

Having implemented the required additions as described in Section \ref{part:userguide}, We shall modify an existing \rivet analysis corresponding to the ATLAS 2010 inclusive jet measurement\cite{Aad:2011fc}, double differential in the rapidity and \pT of the hardest jet in the event. For the purposes of demonstration we shall consider only the lowest rapidity bin of the $R=0.4$ measurement.

As in this section we aim to demonstrate the reproduction of the cross section up to the available interpolation accuracy, we shall make use of the ability of \packagename to base multiple grids with different settings on the same \rivet histogram instance. Our grids are initialised in the \lstinline[language=bash]{init} phase as so:
\clearpage
\begin{lstlisting}[language=c++]
	// Common grid properties
	const string PDFname = "basic.config";
	const double xmin 	= 1E-5;
	const double xmax 	= 1;
	const double Q2min 	= 20;
	const double Q2max 	= 1E7;
	const int LOpower 	= 2;

	// Book low precision interpolation grid
	lowPrec =  MCgrid::bookGrid(pTHisto_R04_y1, histoDir(),
							PDFname, LOpower, xmin, xmax, 
							Q2min, Q2max,
							MCgrid::gridArch(30,20,5,5)   ) );
	
	// Book medium precision interpolation grid
	medPrec =  MCgrid::bookGrid(pTHisto_R04_y1, histoDir(),
							PDFname, LOpower, xmin, xmax, 
							Q2min, Q2max,
							MCgrid::gridArch(40,25,5,5)   ) );

	// Book high precision interpolation grid
	highPrec =  MCgrid::bookGrid(pTHisto_R04_y1, histoDir(),
							PDFname, LOpower, xmin, xmax, 
							Q2min, Q2max,
							MCgrid::gridArch(50,30,5,5)   ) );
\end{lstlisting}
The three booked grids differ only by the content of their \lstinline[language=c++]{gridArch} struct which determines their interpolation accuracy. These grids are then filled after experimental cuts as described in section \ref{part:userguide} alongside the corresponding histogram. Running this rivet analysis over a sample of NLO Dijet events generated by \Sherpa\!+{\tt BlackHat} for the phase space and fill runs, \packagename outputs three \appl {\tt ROOT} format files.

\begin{figure}[ht!]
\centering
\includegraphics[width=0.9\textwidth]{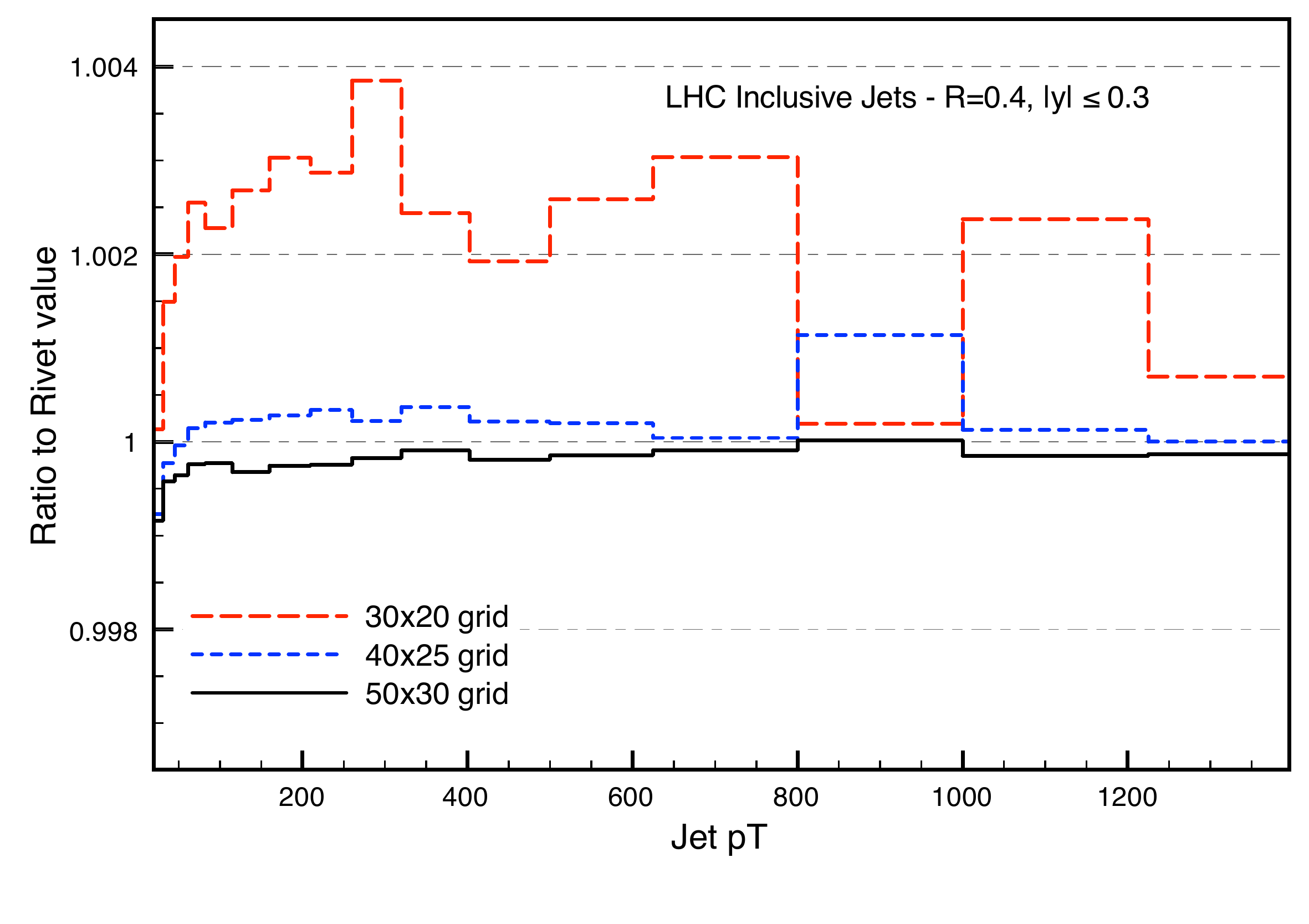}
\caption{Comparison of the reproduction accuracy for three interpolation grid choices for the ATLAS jet production analysis, see text for details. The curves show the ratio of the \appl result to the original \rivet calculation.  }
\label{fig:gridprec}
\end{figure}

As in this analysis we have used the {\tt basic_pdf} subprocess PDF which contains all 121 partonic channels as independent subprocesses, a direct comparison to the original \rivet calculation is possible. In Figure \ref{fig:gridprec} we show the result of convoluting the three produced \appl files, produced from 10 million events, with the generating PDF (in this case, the CT10 NLO set \cite{Lai:2010vv}) and taking the ratio to the benchmark \rivet result.

The results demonstrate that the \packagename interface is able to convert event weights from the Monte Carlo sample into \appl fill calls without any loss of accuracy at the precision of the \appl interpolation result. In all three cases reproduction is at the permille level or better. As detailed above, the accuracy of the \appl calculation may be tuned through the interface to the user's requirements. For a comprehensive discussion of the interpolation accuracy of the \appl framework please refer to \cite{Carli:2010rw}.

\subsection{Parameter variation and the CDF Z rapidity measurement.}
\label{sec:DYvalid}
Having verified that the \packagename interface is able to reproduce the benchmark distribution available in \Sherpa up to the interpolation accuracy available in \appl\!, we shall now verify the ability of the interface to actually reweight the produced event sample with different PDFs and perturbative scales. With this investigation we shall also demonstrate the interface with a different process at a different collider, modifying the existing \rivet analysis for the measurement of the $Z$ boson rapidity distribution at CDF\cite{Aaltonen:2010zza}. To test our modifications we generated two Drell-Yan event samples at NLO in QCD with \Sherpa\!, one with 10 million and one with 100 million events.  The renormalisation and factorisation scales we fixed to $\mu_F^2=\mu_R^2 = M_Z^2$. Both samples we ran through the \packagename enabled analysis. Additionally the event generation was repeated with the renormalisation and factorisation scales varied to $\mu_F^2=\mu_R^2 = 2M_Z^2$ and  $\mu^2_F=\mu^2_R = M_Z^2/2$ such that we can test the reproduction of these runs with the scale variation formula described in Ref.  \cite{Carli:2010rw}. 
\begin{figure}[ht!]
\centering
\begin{subfigure}[a]{0.5\textwidth}
\includegraphics[width=1\textwidth]{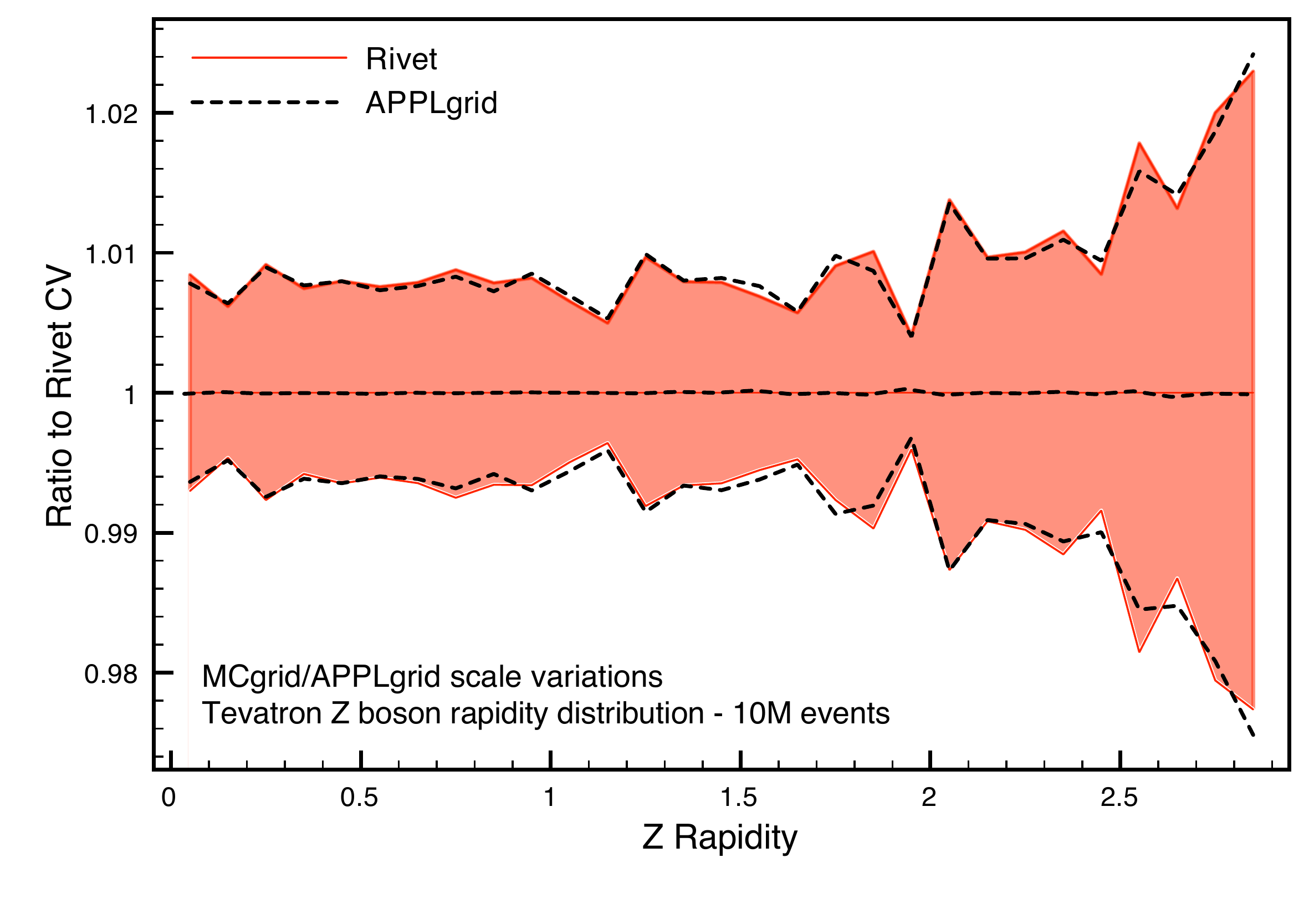}
\caption{10M event sample}
\end{subfigure}\begin{subfigure}[a]{0.5\textwidth}
\includegraphics[width=1\textwidth]{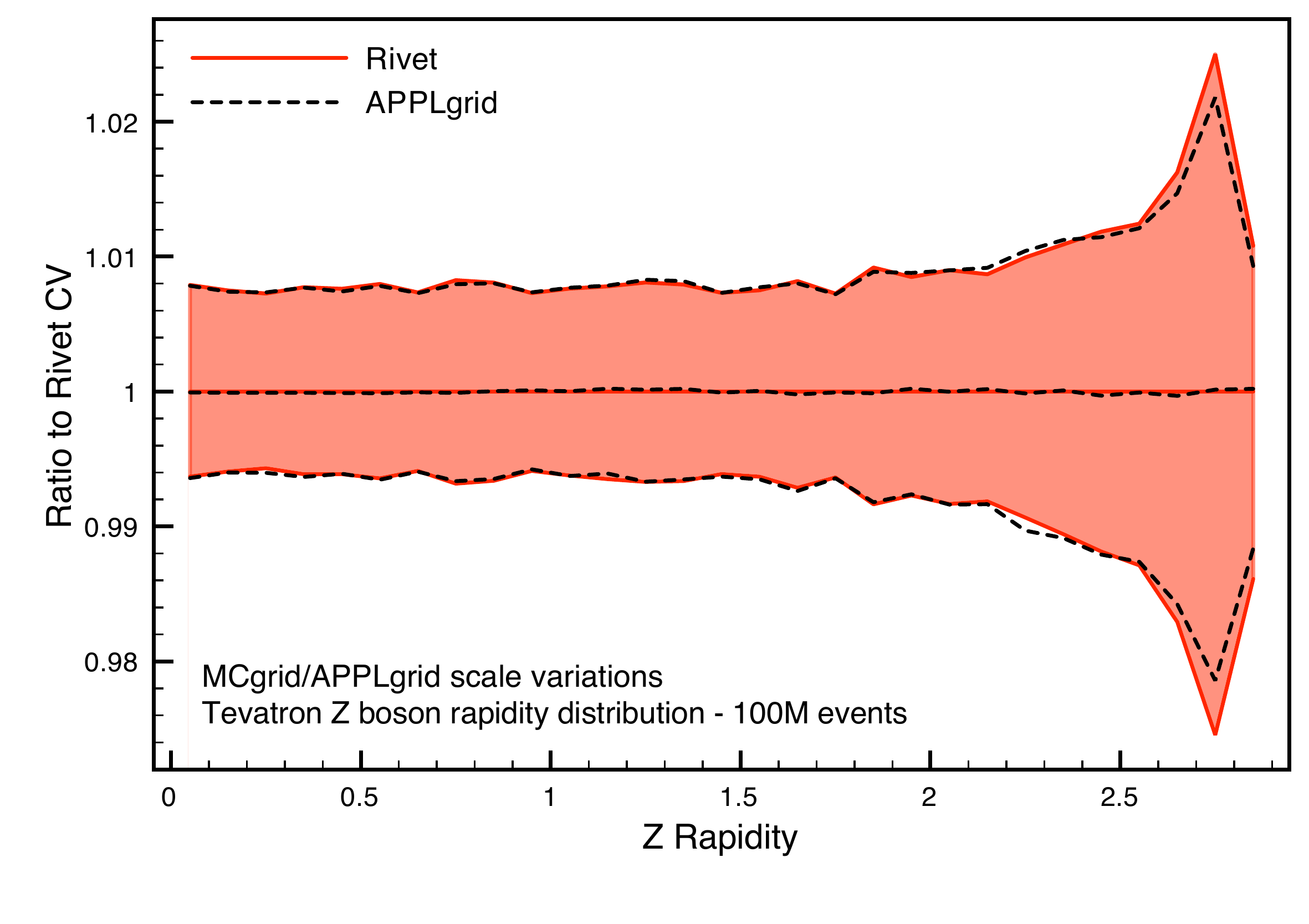}
\caption{100M event sample}
\end{subfigure}
\caption{Comparison of \Sherpa\!/\rivet result to predictions from \appl\!. Plots shown demonstrate the central value and scale variation reproduction for two event samples, with 10 million and 100 million events respectively. The red band represents the \rivet prediction and scale variation uncertainty, and the dotted black line the \appl prediction. All points are normalised to the \rivet central value.}
\label{fig:scalevar1}
\end{figure}

\begin{figure}[ht!]
\centering
\begin{subfigure}[a]{0.5\textwidth}
\includegraphics[width=1\textwidth]{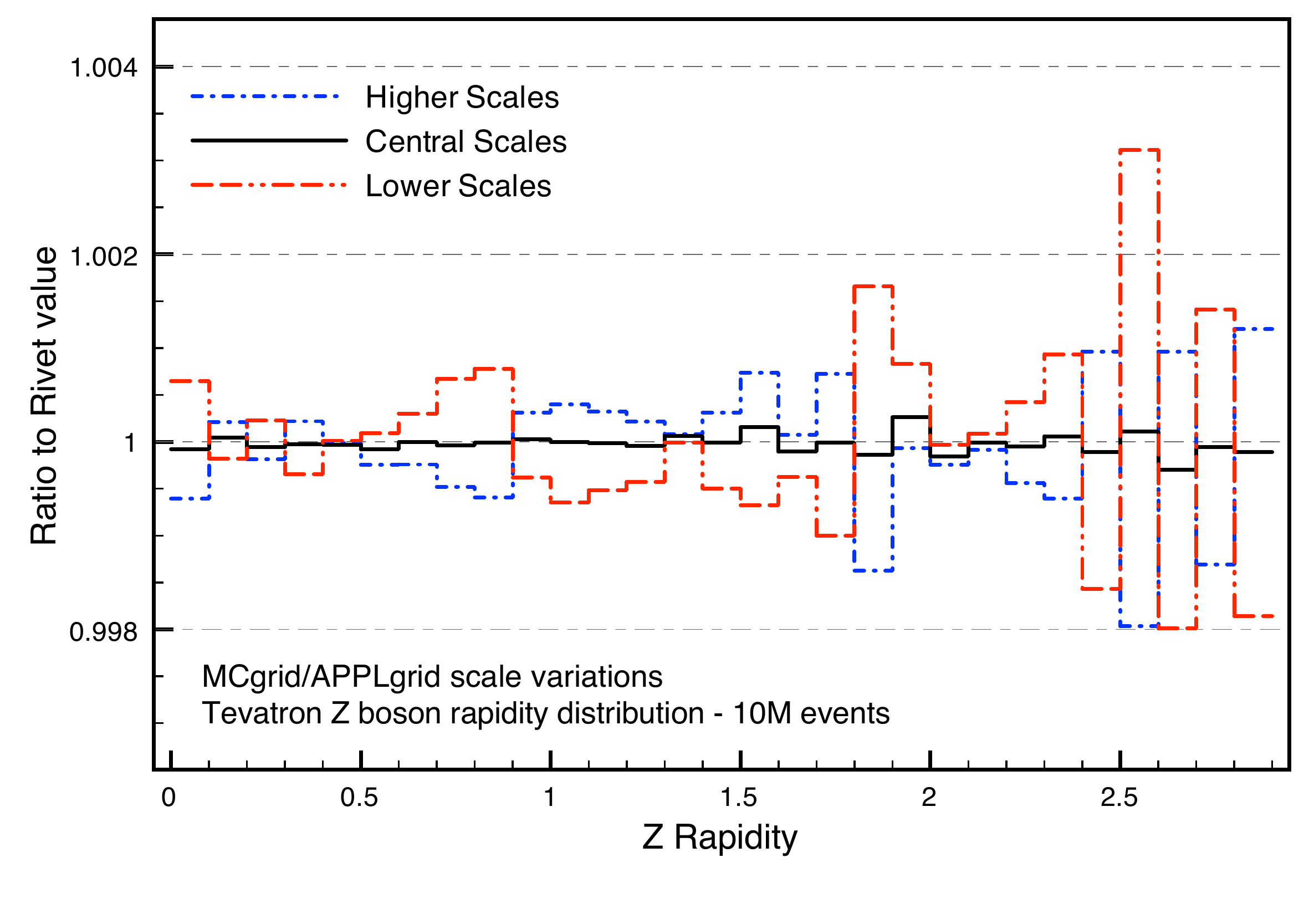} \caption{10M event sample}
\end{subfigure}\begin{subfigure}[a]{0.5\textwidth}
\includegraphics[width=1\textwidth]{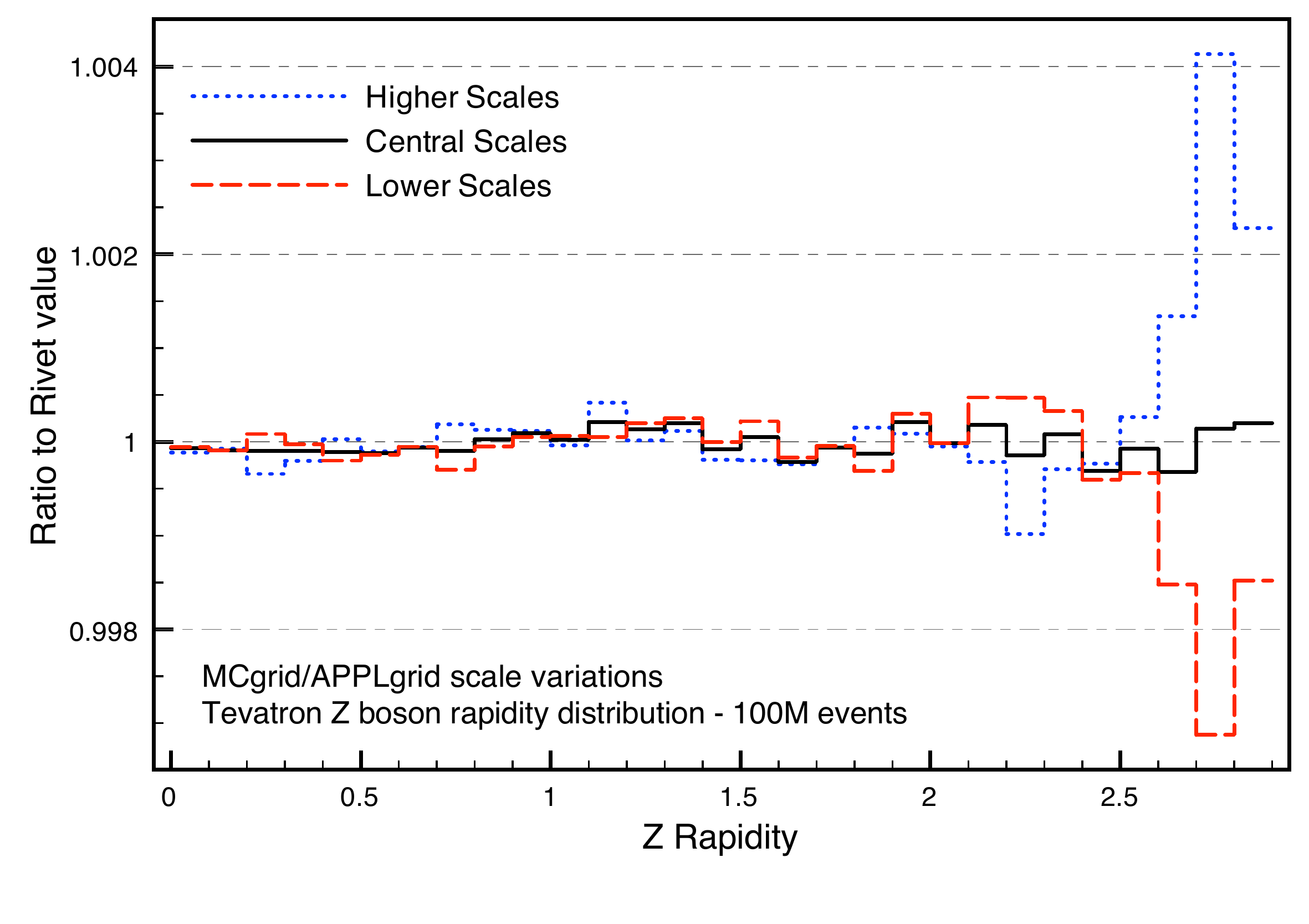}\caption{100M event sample}
\end{subfigure}
\caption{Comparison of \Sherpa\!/\rivet result to predictions from \appl\!. Plots shown demonstrate the central value and scale variation reproduction for two event samples, with 10 million and 100 million events respectively. Each histogram represents the ratio of the \appl result to its equivalent result from \Sherpa\!/\rivet.}
\label{fig:scalevar2}
\end{figure}

While the reproduction of the central scale distribution is limited only by interpolation accuracy, the reproduction of the scale variations does have an additional sensitivity to the statistical accuracy of the grid, as the accurate determination of the missing logarithmic terms in the fill weights depends on the interpolation grids being well populated. In particular the accurate reproduction of the scale variation uncertainties is a nontrivial test not only of the \appl interpolation but also of the correct weight conversion and PDF dependence removal performed by \packagename.

In Figures \ref{fig:scalevar1} and \ref{fig:scalevar2} the agreement between the interpolated predictions and the benchmark result is demonstrated in the case of central values and scale variations for the two event samples. For both samples, the reproduction of the central value is excellent, with precision only limited by the interpolation accuracy. The reproduction of the scale variations is also very good, with accuracy typically much better than the percent level. The effect of increasing statistics is clear also in the improvement between the two samples for the varied scales.  This provides a strong validation of the \appl scale variation formula for Monte Carlo weights arising from a fully exclusive event generator.

\begin{figure}[ht!]
\centering
\includegraphics[width=0.5\textwidth]{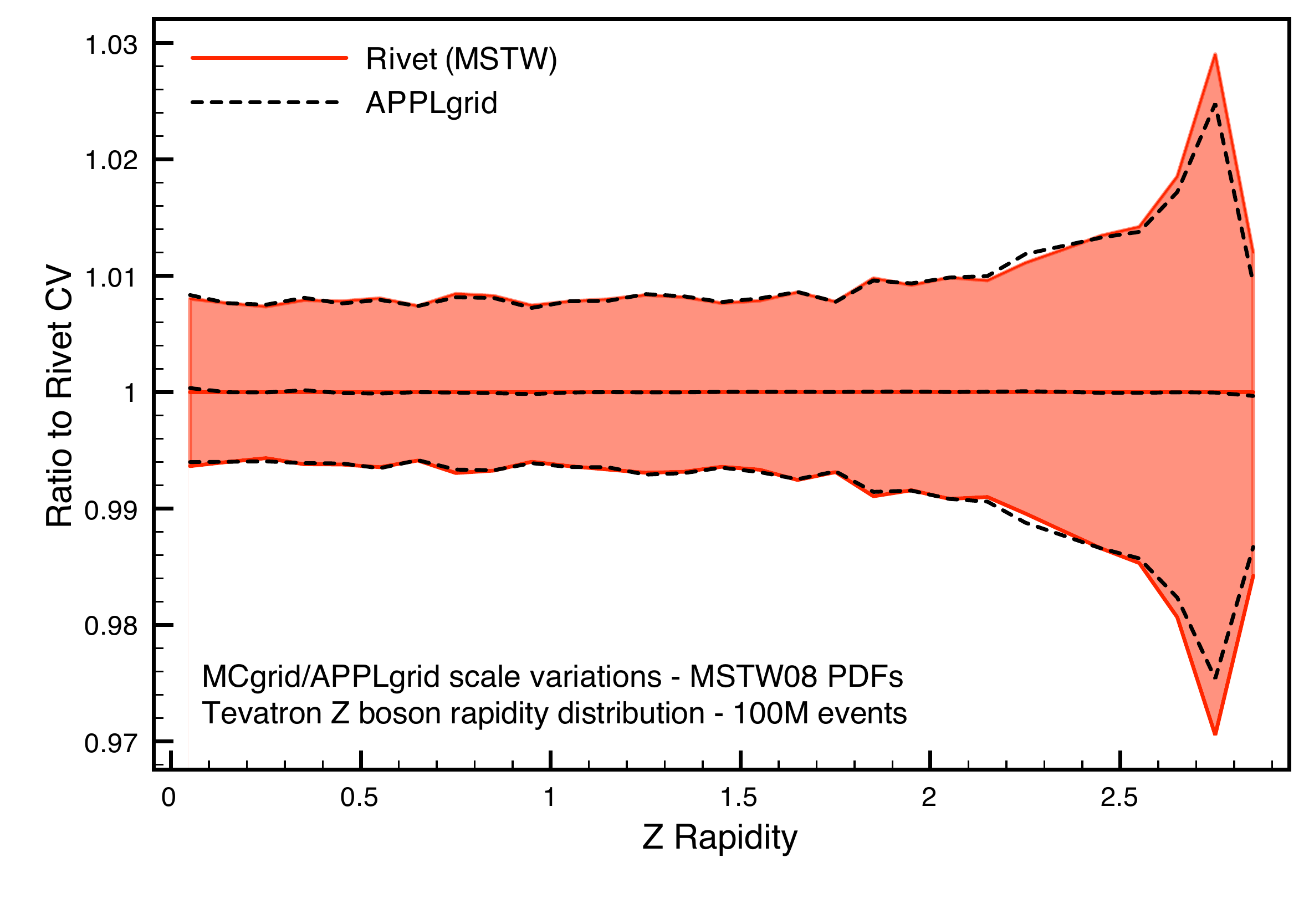}\includegraphics[width=0.5\textwidth]{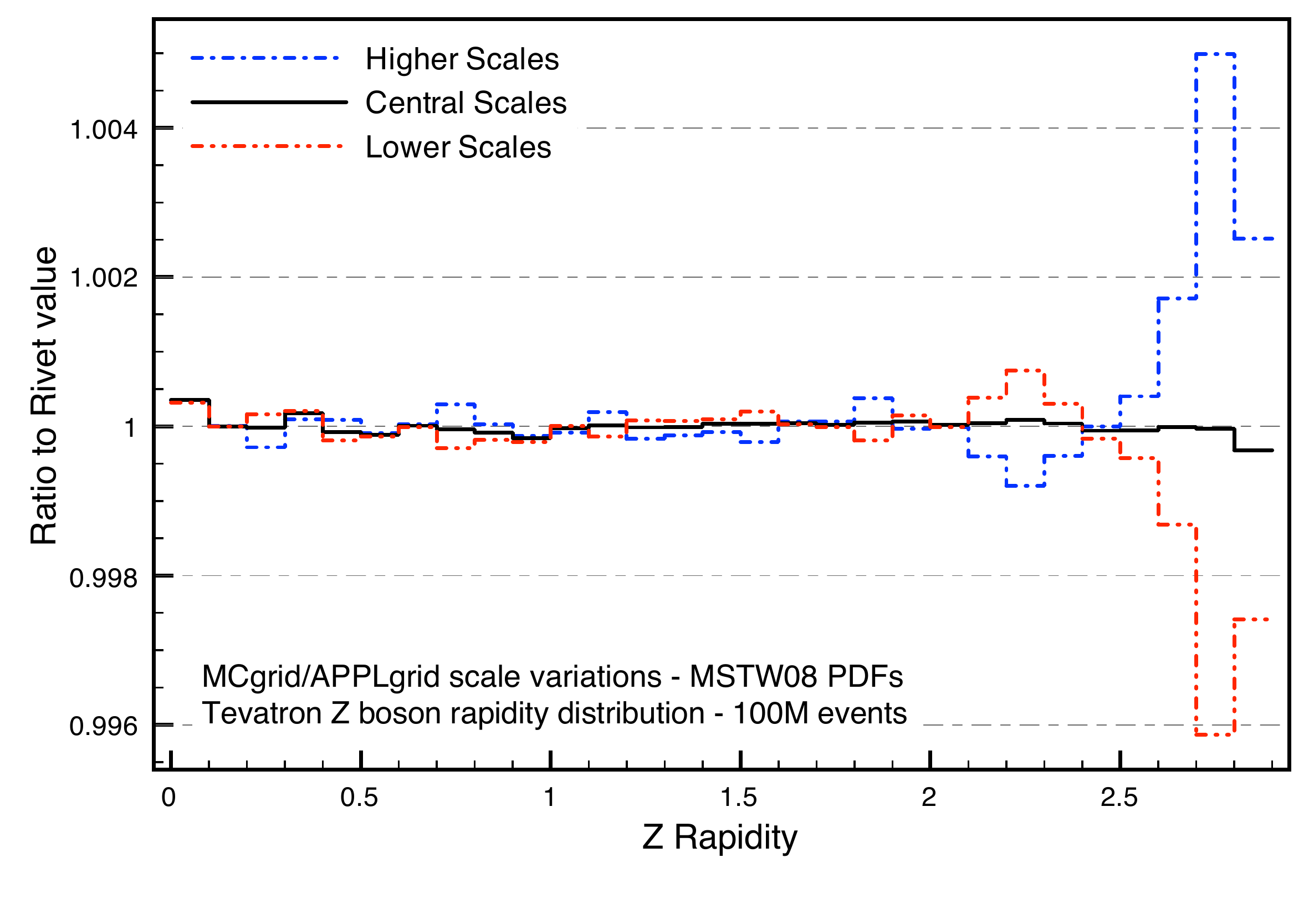}
\caption{Comparison of \Sherpa\!/\rivet result using MSTW08 PDFs with predictions from an \appl generated with CT10 and convoluted with MSTW08 PDFs. Plots shown demonstrate the central value and scale variation reproduction for the 100 million event sample grid. The red band in the left panel represents the \rivet prediction and scale variation uncertainty, and the dotted black line the \appl prediction. All points are normalised to the \rivet (MSTW) central value.}
\label{fig:mstwDY}

\end{figure}

The event generation has also been performed under a variation of the PDF used in the calculation. To verify the produced grids under their most typical application, PDF reweighting, the \Sherpa\!/\rivet run was repeated with the use of MSTW2008 PDFs~\cite{Martin:2009iq}. Using the original grid generated with the CT10 distributions, we perform the product with MSTW2008 to test if the produced grids are able to reweight PDFs effectively. The results as demonstrated in Figure \ref{fig:mstwDY} show that the PDF dependence of the event weights has been properly removed, allowing for the correct reweighting when re-convoluted with a different PDF set.  

\subsection{Grid fills with subprocess identification}
As described in Section \ref{sec:subprocstat}, the use of a subprocess basis for the incoming PDFs makes a direct comparison to the \rivet result more complicated, as the statistical accuracy of the result is modified. In addition, the \packagename interface also must perform a tracking of the relative statistical population of the individual partonic channel contributions to a subprocess, such that their selection weights may be corrected to the subprocess' weight.

\begin{figure}[ht!]
\centering
\begin{subfigure}[a]{0.5\textwidth}
\includegraphics[width=1\textwidth]{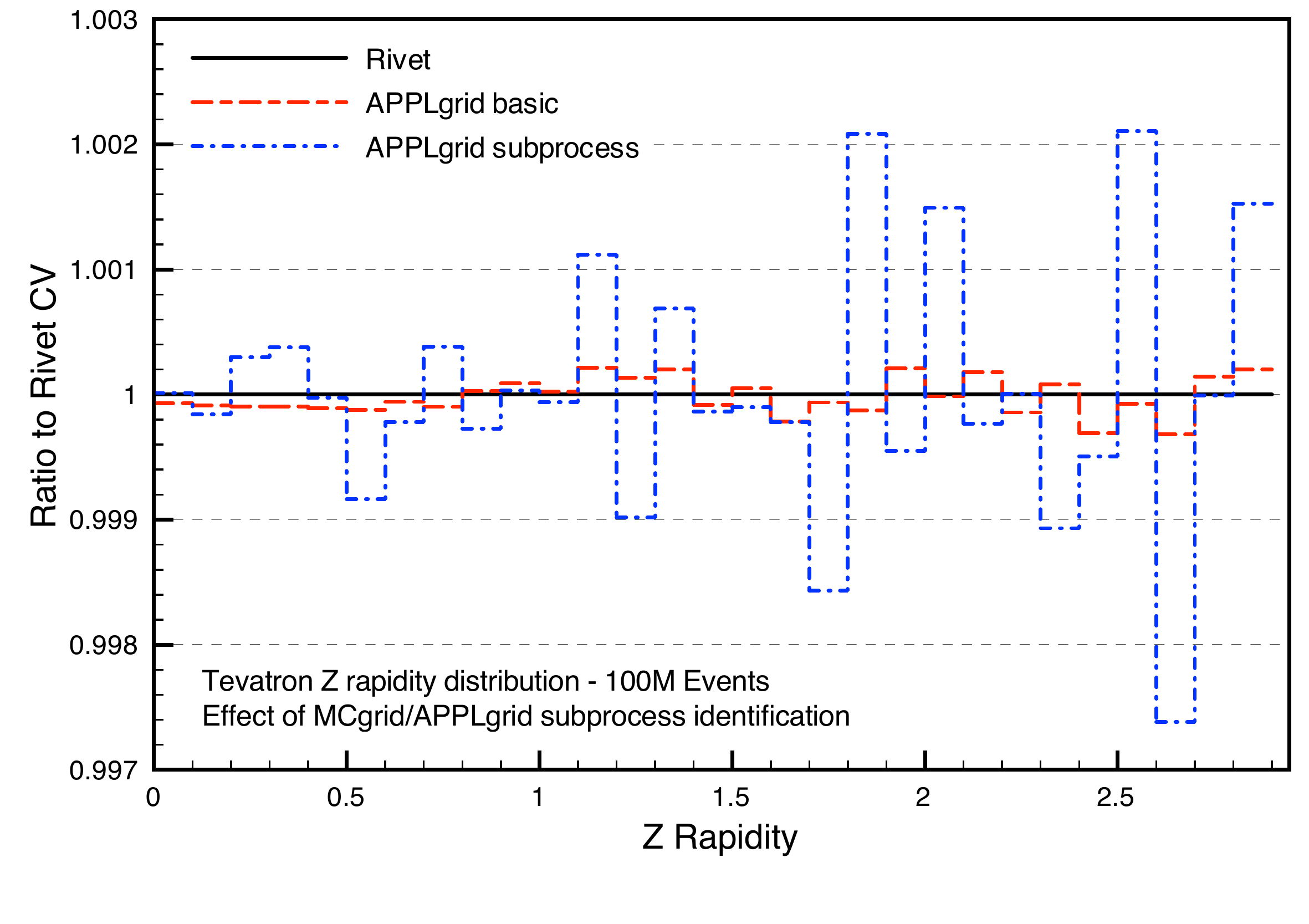} \caption{Ratio to \rivet result.}
\end{subfigure}\begin{subfigure}[a]{0.5\textwidth}
\includegraphics[width=1\textwidth]{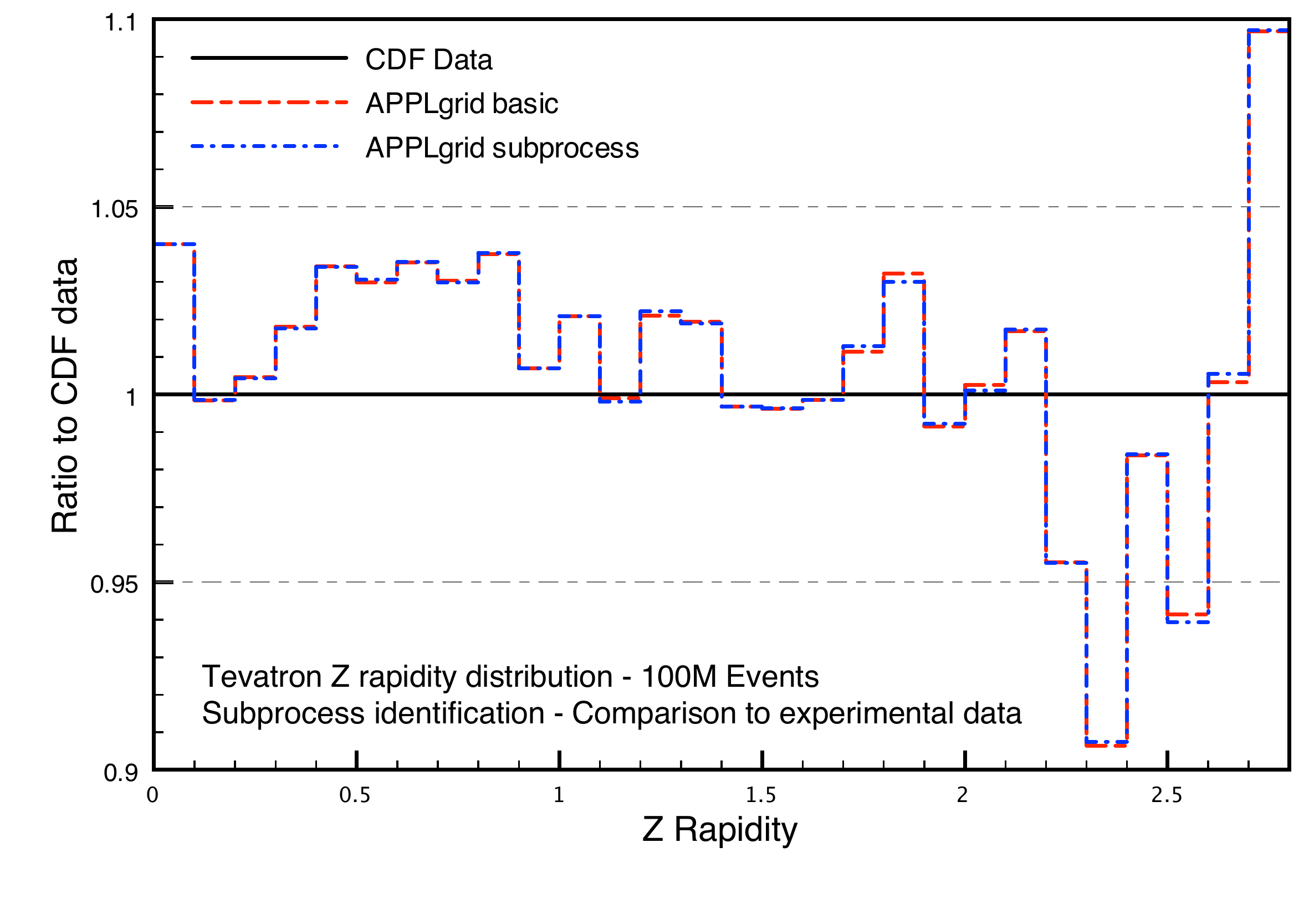}\caption{Ratio to CDF data.}
\end{subfigure}
\caption{Illustration of the typical differences expected between produced grid files utilising either the {\tt basic} set of subprocesses or with the subprocesses identified through the scripts included in the \packagename package. The left plot shows the deviation from the benchmark \rivet result, and the right demonstrates that the deviation does not correspond to a reduced level of agreement with experimental data.}
\label{fig:subprocAccuracy}
\end{figure}

To test the \packagename implementation of this tracking, and its generation of subprocess-identified \appl files, we performed the 100M Drell-Yan fill as described in section \ref{sec:DYvalid} but this time utilising the {\tt lumi_pdf} config file generated by the packaged scripts. In Figure \ref{fig:subprocAccuracy} the ratio of the \appl result to the benchmark distribution is shown, with the {\tt basic_pdf} result from Section \ref{sec:DYvalid} for comparison.  It is important to note that the reduced agreement between the subprocess \appl and the \rivet benchmark does not imply that the subprocess result is less accurate, rather that here we are not directly comparing like with like. 

\subsection{Demonstration of \appl convolutions}
Finally, to provide a demonstration of the performance of the \appl interpolation methods and illustrate the potential applications of grids produced via the \packagename interface, we shall examine two cases where a large number of repeat calculations must be performed.

Using the grid files produced in section \ref{part:validation} for the validation of the interface,
the uncertainty from perturbative scale variations and $\alpha_s$ may be assessed in a nonlinear fashion by using the representation of probability distributions in the space of PDFs available in NNPDF 2.3~\cite{Ball:2012cx}. 

\begin{figure}[ht!]
\centering
\begin{subfigure}[a]{0.5\textwidth} 
\includegraphics[width=1\textwidth]{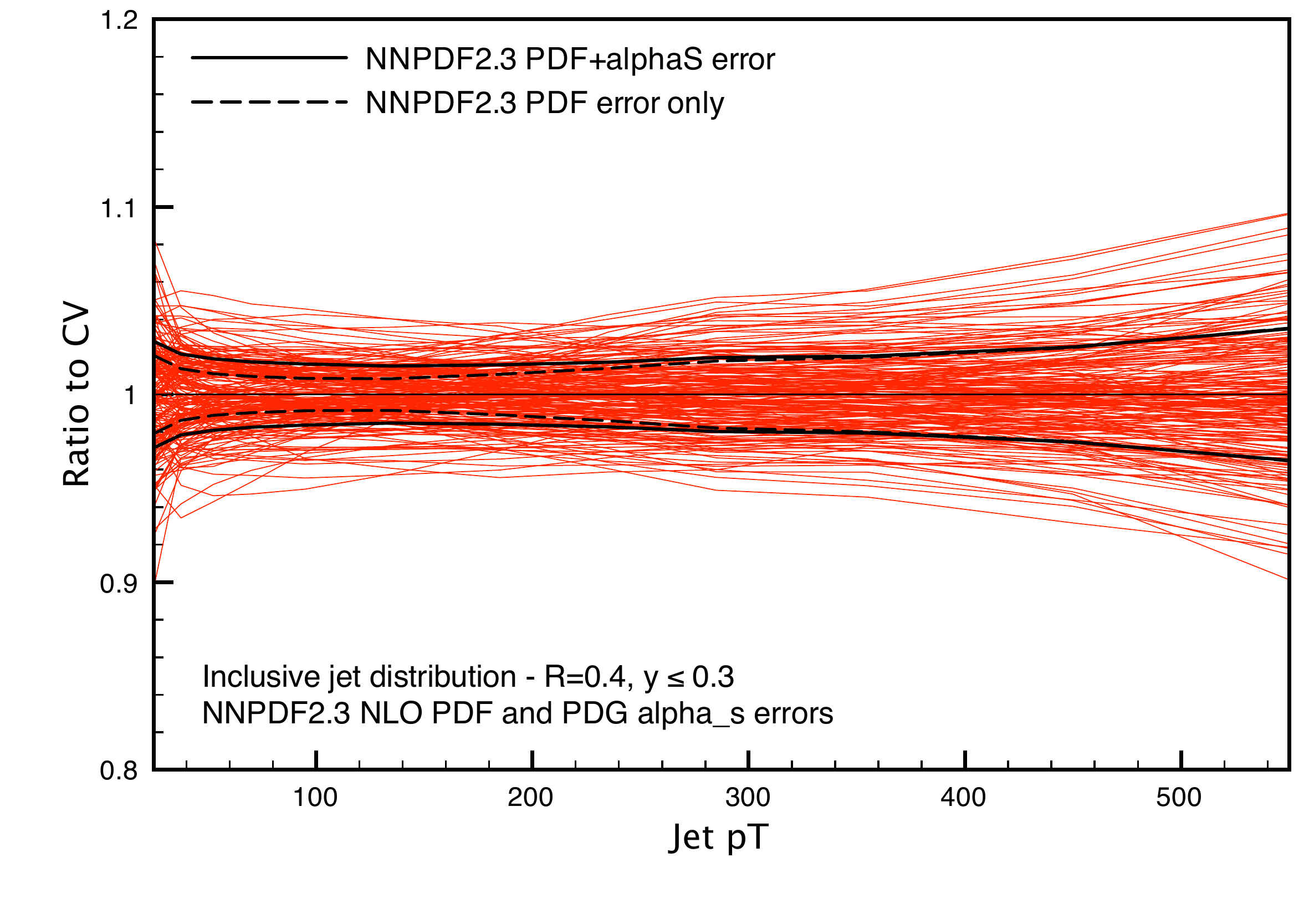} \end{subfigure}\begin{subfigure}[a]{0.5\textwidth}

\includegraphics[width=1\textwidth]{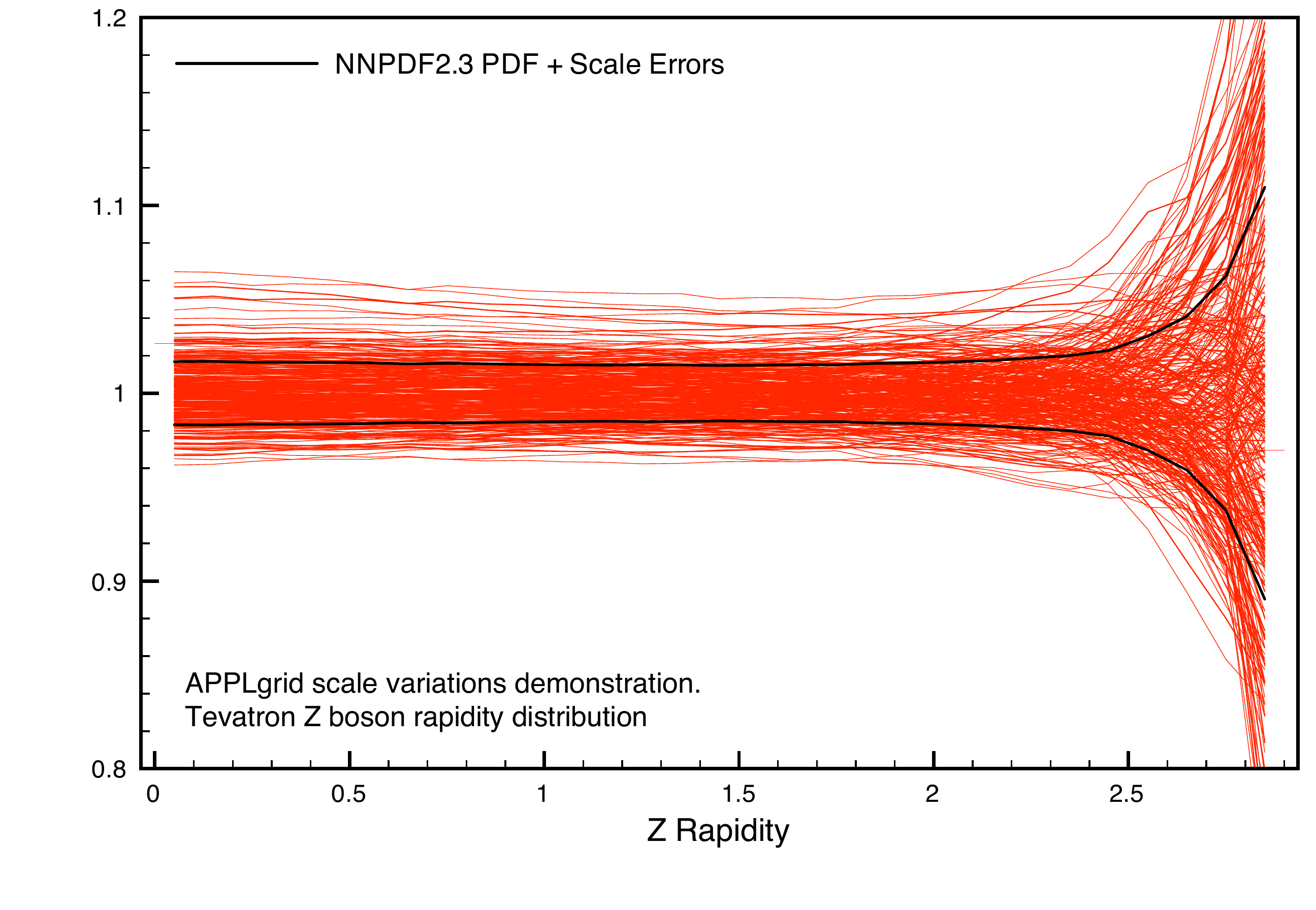}\end{subfigure}
\caption{Example application of the \packagename produced \appl files. The left figure shows the NNPDF2.3 replica distribution for an inclusive jet \pT distribution at the LHC, including the error on $\alpha_s$ via the replica distribution. The figure on the right shows predictions for the CDF Z rapidity measurement, with replica distributions for the central scale, and variations. On each plot, the red lines show individual NNPDF2.3 replicas, the black lines denoting the 1-$\sigma$ contours.}
\label{fig:examples}
\end{figure}

Taking the PDG reference value of $\alpha_s=0.1184(7)$~\cite{Beringer:1900zz}, it is possible to generate a distribution of PDFs for this value including the uncertainty by sampling appropriately from the available NNPDF2.3 $\alpha_s$ sets. Assuming a Gaussian uncertainty around $\alpha_s=0.1184$ and normalising to the maximum number of PDF replicas available in each set, this corresponds to taking 16 replicas at $\alpha_s=0.117$, 100 at $\alpha_s=0.118$, 82 at $\alpha_s=0.119$ and finally 9 replicas at $\alpha_s=0.120$.
Using this total sample of 207 PDF replicas, we plot the combined PDF and $\alpha_s$ uncertainty for the inclusive jet pT distribution in figure \ref{fig:examples}.

Going to an even larger basis of PDF replicas, In the right of figure \ref{fig:examples} we show the replica distribution for the NNPDF2.3 CDF Z rapidity prediction including scale variation uncertainties. In this instance there are a total of 300 predictions computed and plotted.

Plotting distributions such as these, with very large numbers of predictions leading to accurate assessments of the underlying uncertainties, is only feasible with the use of interpolating tools such as \appl. The \packagename interface now greatly increases the number of  processes available for such interpolation.

%% file: conclusions.tex
In this paper we have presented \packagename\!, a package for producing
\appl tables from samples of events produced by Monte Carlo
generators. These tables are based on interpolating functions that
allow for precise, fast, and flexible computations of scale variations
and PDF reweighting. In this way, the storage requirements and the
speed of these calculations is greatly improved. 

\packagename provides additional methods to be used in conjunction with
the \rivet programme. In the analysis of a Monte Carlo calculation of 
a fixed order process it allows for the production of an \appl for every considered 
observable.
 
The basic idea follows the general \appl prescription, whereby grids
are computed using some interpolating functions for the PDFs, and
separating explicitly the dependence on the perturbative order, the
renormalisation and factorisation scales as discussed in
Sect.~\ref{part:interface}. Note that these interpolation tables are
computed by summing over the generated events, for a specific choice
of the kinematical variables. The differential cross section is then
obtained by contracting these tables with the values of the PDFs at
the points chosen in the interpolation grids. The choice of the
interpolation grids determines the accuracy of the interpolation
tool. Once again, the structure of the interpolating grids has to be
decided in advance, and cannot be changed once the tables have been
produced.

Using the interpolation tables at LO in perturbation theory is a
straightforward exercise. The extension of the tool at NLO is more
subtle, because the precise PDF dependence of the integrated subtraction
terms must be taken properly into account. We have detailed our NLO
implementation in Sect.~\ref{part:interface}. 

The details of the software implementation have been presented in
Sect.~\ref{part:userguide}. 

The interface has been validated by studying two processes, namely the
inclusive jet production at the LHC, and the Drell-Yan production of $Z$
bosons at the Tevatron. As expected, the grids can be tuned to reach an 
excellent accuracy of order $10^{-3}$ for the computation of the
observables. Using the grids we provide an explicit example of the
parameter variations mentioned above. 

\packagename enables faster studies of scale variations, and PDF
variations, without having to perform multiple runs of the Monte Carlo
generators. This allows for the determination of reliable uncertainty estimates for 
arbitrary observables even for very complicated and computationally 
challenging multi-particle final state calculations. It also provides a 
solution to the large storage requirements that are necessary for other 
methods relying on storing explicit events~\cite{Bern:2013zja}. Because 
of the increased performance in computing observables, \packagename paves 
the way for the inclusion of more observables in modern PDF fits. In its present implementation, the full PDF dependence of showered and hadronised events
is not fully accounted for. This would require tracing the PDF dependence of the parton 
shower history of individual events, which is beyond the scope of this 
publication. While \packagename is able to process such events into the \appl format, 
there is an implicit approximation present in that the reweighting is only performed at the level of
the hard process. Therefore full NLO accuracy can only be claimed for fixed order calculations.

\packagename is publicly available and can be downloaded from 
\url{http://mcgrid.hepforge.org}. We make use of the {\tt HepMC} 
event record, requiring some additional event 
information being stored in the {\tt HepMC::WeightContainer}. For 
the {\Sherpa} event generator, as of version {\tt 2.0.0}, this 
information is provided by default with the {\tt HepMC_Short} output 
format.

%% file: acknowledgements.tex
We wish to thank the \appl developers Tancredi Carli, Pavel Starovoitov and 
Mark Sutton for fruitful discussions and help provided. Furthermore we are grateful 
for support on {\tt Rivet} and \Sherpa from Frank Siegert, Stefan H\"oche and Marek Sch\"onherr. 
We want to thank Enrico Bothmann for extensive tests of the code and Alberto Guffanti for advice and comments.
We acknowledge support from the EU MCnetITN research network. MCnetITN is a Marie Curie Training Network funded 
under Framework Programme 7 contract PITN-GA-2012-315877.